\documentclass[aps,pra,twocolumn,floats,showpacs,10pt,a4paper]{revtex4-1}

\usepackage{bbm} 
\usepackage{amssymb} 
\usepackage{ifpdf} 
\usepackage{color}
\usepackage{graphicx,epsfig}
\usepackage{amsmath} %align environment 
\usepackage{wasysym} %Hexagon character
\usepackage{todonotes}
\usepackage{natbib}
\usepackage{hyperref}
\newcommand{\Z}{{\mathbb{Z}}} \newcommand{\R}{{\mathbb{R}}}
 \newcommand{\C}{{\mathbb{C}}}

\begin{document}

\title{From the $SU(2)$ Quantum Link Model on the Honeycomb Lattice \\ to the Quantum Dimer Model on the Kagom\'e Lattice: \\ Phase Transition and Fractionalized Flux Strings}

\author{
D.\ Banerjee$^1$, 
F.-J.\ Jiang$^2$, 
T.\ Z.\ Olesen$^3$, 
P.\ Orland$^4$, 
and U.-J.\ Wiese$^3$
}
\vspace{10pt}
\affiliation{
$^1$ NIC, DESY, Platanenallee 6, D-15738 Zeuthen, Germany \\ 
$^2$ Department of Physics, National Taiwan Normal University 88, Sec.\ 4, Ting-Chou Rd., Taipei 116, Taiwan \\ 
$^3$Albert Einstein Center for Fundamental Physics, Institute for Theoretical Physics, University of Bern, Switzerland \\ 
$^4$ Baruch College, The City University of New York, 17 Lexington Avenue, New York, NY 10010, USA, and \\
Graduate School and University Center, The City University of New York, 365 Fifth Avenue, New York, NY 10016, USA
}

\begin{abstract}
We consider the $(2+1)$-d $SU(2)$ quantum link model on the honeycomb lattice and show that it is equivalent to a quantum dimer model on the Kagom\'e lattice. The model has crystalline confined phases with spontaneously broken translation invariance associated with pinwheel order, which is investigated with either a Metropolis or an efficient cluster algorithm. External half-integer non-Abelian charges (which transform non-trivially under the $\Z(2)$ center of the $SU(2)$ gauge group) are confined to each other by fractionalized strings with a delocalized $\Z(2)$ flux. The strands of the fractionalized flux strings are domain walls that separate distinct pinwheel phases. A second-order phase transition in the 3-d Ising universality class separates two confining phases; one with correlated pinwheel orientations, and the other with uncorrelated pinwheel orientations.

\end{abstract}

\maketitle

\section{Introduction}
\label{SEC:Introduction}

Quantum link models provide a generalization of Wil\-son-ty\-pe lattice gauge theories, in which the link variables are not classical parallel transporters but intrinsically quantum mechanical objects, similar to generalized quantum spins. The first quantum link models with gauge groups $U(1)$ and $SU(2)$ were formulated by Horn in 1981 \cite{Hor81}. These models were studied in more detail by Orland and Rohrlich under the name of gauge magnets \cite{Orl90}. In \cite{Cha97} quantum link models were used as an alternative regularization of non-Abelian gauge theories. Quantum link models with an $SU(N)$ gauge symmetry were constructed in \cite{Bro99} and were introduced as an alternative formulation of lattice Quantum Chromodynamics (QCD). In this formulation, 4-d continuous gluon fields emerge via dimensional reduction from the collective dynamics of $(4+1)$-d discrete quantum link variables, and quarks manifest themselves as domain wall fermions at the edge of the extra dimension.

Thanks to their finite-dimensional Hilbert space per link, quantum link models are well suited for quantum simulation of dynamical gauge theories with ultracold matter. In particular, the discrete quantum link degrees of freedom can be embodied by a few quantum states of ultracold atoms in an optical lattice \cite{PhysRevLett.109.175302,PhysRevLett.110.125303,ANDP:ANDP201300104}. Although the ultimate long-term goal is to quantum simulate QCD in order to address its real-time evolution as well as its phases at non-zero baryon density, quantum simulation experiments will have to start with much simpler toy-model gauge theories. One of the simplest models is the $U(1)$ quantum link model in which a single quantum spin 1/2 per link represents the gauge degrees of freedom. Quantum simulators for this model have been proposed using ultracold Rydberg atoms in optical lattices \cite{PhysRevX.4.041037}, or alternatively systems of superconducting flux circuits \cite{MARCOS2014634}.

The $(2+1)$-d $U(1)$ quantum link model has been si\-mu\-la\-ted with an efficient cluster algorithm (applied in Euclidean time using quantum Monte Carlo simulations on a classical computer) \cite{Ban13a}. Interestingly, the model has two distinct confined phases, separated by a rather weak first-order phase transition, which ``masquerades'' as a deconfined quantum critical point \cite{Senthil1490}. Both phases spontaneously break translation invariance by one lattice spacing, and thus give rise to ``crystalline confinement''. In one of the two phases, in addition, charge conjugation is spontaneously broken. In both phases, the confining electric flux string which connects an external static charge with an anti-charge, fractionalizes into different strands, each carrying 1/2 unit of electric flux. The strands play the role of domain walls separating the two $\Z(2)$ realizations of a given type of confined phase, which coexist due to spontaneous translation symmetry breaking. The interior of these strands has a remarkable feature: it consists of the other type of confined phase (which exists in the bulk on the other side of the phase transition) \cite{Ban13a}.

The $(2+1)$-d $U(1)$ quantum link model on the square lattice also has interesting connections to condensed matter physics. In particular, it has the same Hamiltonian as the square lattice quantum dimer model \cite{PhysRevLett.61.2376,PhysRevB.65.024504,PhysRevB.81.214413}, which, however, realizes a modified Gauss law with staggered background charges. Again, using quantum Monte Carlo, the controversially discussed phase structure of the square lattice quantum dimer model has been clarified \cite{PhysRevB.94.115120}. It was found that the columnar phase extends all the way to the so-called Rokhsar-Kivelson point, without any intervening plaquette or mixed phases. In this case, the confining strings connecting an external charge-anti-charge pair fractionalize into strands that carry even just 1/4 unit of electric flux. In this case, the strands represent domain walls that separate coexisting columnar phases.  Interestingly, their interior consists of plaquette phase, although this phase is not realized in the bulk.

In this paper, we extend the study of $(2+1)$-d quantum link models from $U(1)$ to the non-Abelian gauge group $SU(2)$, which has the center $\Z(2)$.  Interestingly, in its vacuum sector, the $SU(2)$ quantum link model on the honeycomb lattice corresponds to the quantum dimer model on the Kagom\'e lattice \cite{PhysRevLett.62.2405,PhysRevLett.89.137202}. We present numerical results --- partly obtained with an efficient cluster algorithm --- that reveal the phase structure of the model. Again, we find two types of crystalline confined phase with spontaneously broken translation symmetry. From the dimer model perspective, these phases display pinwheel order. In one type of confined phase, the orientation of the pinwheels is correlated over infinite distances. In the other type of confined phase pinwheel order still persists, but the orientations of the pinwheels are no longer correlated. In this case, the phase transition that separates the two types of confined phases is second order and consistent with the universality class of the 3-d Ising model.

In the $SU(2)$ quantum link model, external static charges are specified by an $SU(2)$ representation, which characterizes how the Gauss law is realized at a lattice site $x$. The non-Abelian charges fall into two categories: those that are associated with an integer ``color-spin'' representation of $SU(2)$ transform trivially under the $\Z(2)$ center, and those associated with a half-integer ``color-spin'' (henceforth, the quotes on this word will be dropped) representation carry non-trivial center electric flux. While half-integer external charges are confined by unbreakable strings, integer external charges can be screened by dynamical gauge fields. In particular, we investigate the strings connecting external charges in the color-spin 3/2 representation. Remarkably, the corresponding string again fractionalizes into two strands with delocalized $\Z(2)$ center electric flux. As in the Abelian model, the strands play the role of domain walls separating different realizations of the same type of confined phase.

The rest of the paper is organized as follows. Section \ref{SEC:QLMHoney} addresses the connections between the \hbox{$(2+1)$-d} $SU(2)$ quantum link model on the honeycomb lattice and the quantum dimer model on the Kagom\'e lattice. In particular, we construct the most general $SU(2)$ gauge invariant ring-exchange Hamiltonian associated with elementary hexagons, that respects the lattice symmetries. In Section \ref{SEC:QLMHoney} we also introduce a dual height variable representation of the model, which is used in the numerical simulations discussed in Section \ref{SEC:MonteCarlo}. There we present results about the phase structure and the nature of the confining strings. Finally, Section \ref{SEC:Conclusion} contains our conclusions. The details of the cluster algorithm are discussed in an appendix.

\section{The $SU(2)$ Quantum Link Model on the Honeycomb and the Quantum Dimer Model on the Kagom\'e Lattice}
\label{SEC:QLMHoney}

In this section we construct the $SU(2)$ quantum link model on the honeycomb lattice and relate it to the quantum dimer model on the Kagom\'e lattice. In particular, we construct the most general hexagon-based $SU(2)$ invariant ring-exchange Hamiltonian that respects the lattice symmetries. We also reformulate the model in terms of dual height variables which will be used in the numerical simulations described in Section \ref{SEC:MonteCarlo}.

\subsection{Algebraic Structure of the $SU(2)$ Quantum Link Model}

\begin{figure}
	\begin{center}
		\includegraphics[width=0.4\textwidth]{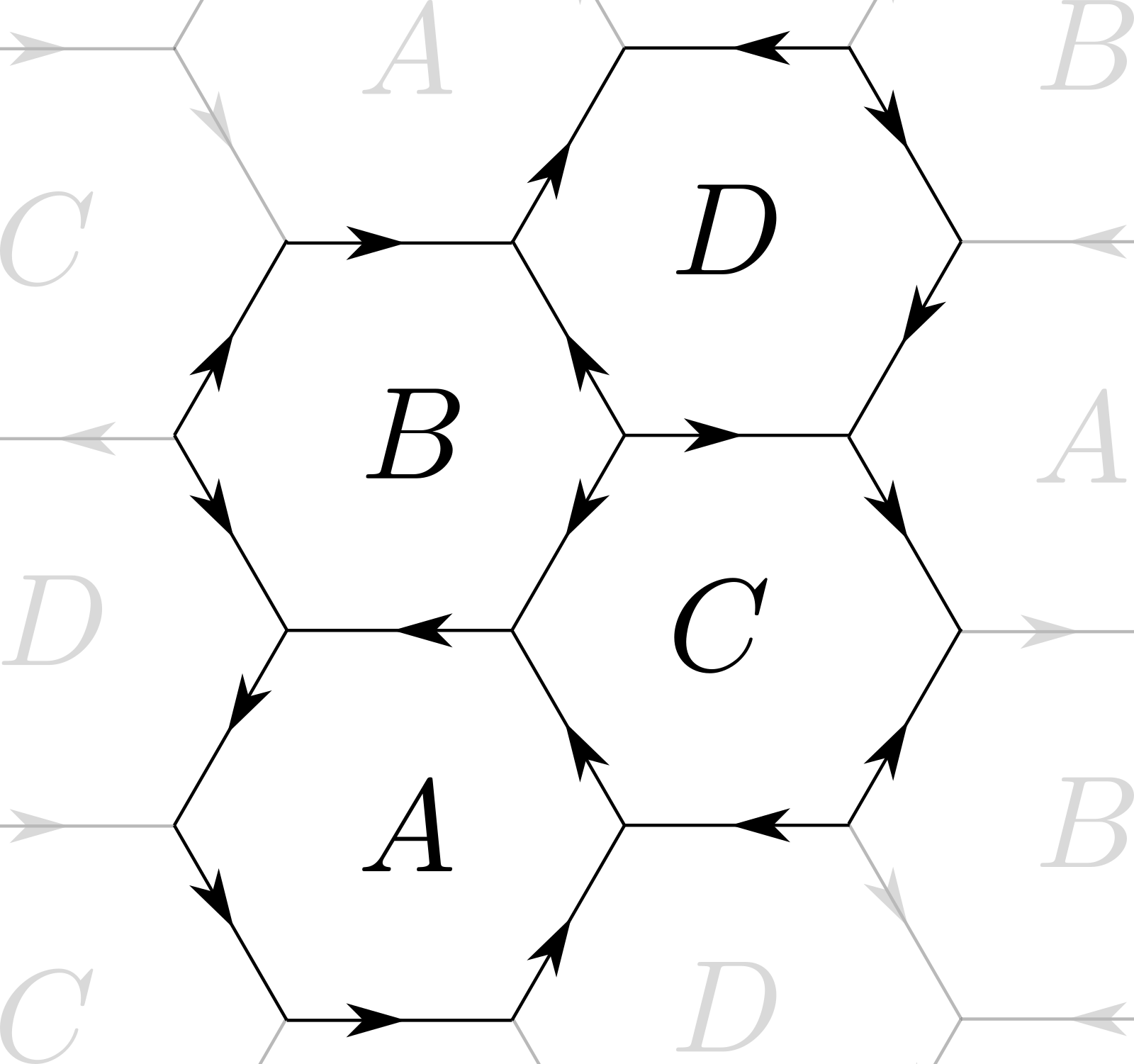}
	\end{center}
	\caption{\it Honeycomb lattice with a suitable choice of link orientations. We distinguish four dual triangular sublattices $A$, $B$, $C$, and $D$, which arise in the context of pinwheel order and of the corresponding height variable representation.
	}
	\label{Honeycomblattice}
\end{figure}

As illustrated in Fig.\,\ref{Honeycomblattice}, we consider a honeycomb lattice with specific link orientations that are chosen in this particular way in order to facilitate the implementation of the Gauss law.

A quantum link operator $U_{xy}$ resides on each of the links connecting neighboring lattice sites $x$ and $y$. Like Wilson's parallel transporters, $SU(2)$ quantum link operators are $2 \times 2$ matrices
\begin{align}
U_{xy} &= U^0_{xy} + i U_{xy}^a \sigma_a, \quad a \in \{1,2,3\},
\end{align}
where $\sigma_a$ are the Pauli matrices. However, the elements of the quantum link matrices are no longer complex numbers but non-commuting operators acting in a finite-dimensional Hilbert space. In particular, $U^0_{xy}$ and $U_{xy}^a$ are represented by four Hermitian operators. Just like Wilson's parallel transporters, under gauge transformations $\Omega_x = \exp(i \omega^a_x \sigma_a) \in SU(2)$ quantum link operators transform as
\begin{align}
\label{gaugeU}
U_{xy}' &= \Omega_x U_{xy} \Omega_y^\dagger = V U_{xy} V^\dagger.
\end{align}
In Hilbert space, gauge transformations are represented by unitary operators
\begin{align}
V &= \prod_x \exp(i \omega_x^a G_x^a).
\end{align}
Here $G_x^a$ is an infinitesimal generator of $SU(2)$ gauge transformations at the site $x$, which obeys
\begin{align}
[G_x^a,G_y^b] &= 2 i \delta_{xy} \varepsilon_{abc} G_x^c.
\end{align}
In accordance with Gauss' law, $G_x^a$ receives contributions from the links connected to the site $x$
\begin{align}
\label{gaugetrafo}
G_x^a &= \sum_y L_{xy}^a + \sum_z R_{zx}^a.
\end{align}
Here the sum over $y$ extends over those nearest neighbors of $x$ for which the connecting link is oriented from $x$ to $y$. The sum over $z$, on the other hand, extends over the nearest neighbors of $x$ for which the connecting link is oriented from $z$ to $x$. The Hermitian operators $L_{xy}^a$ and $R_{xy}^a$ generate $SU(2)$ gauge transformations at the ``left'' ($x$) and the ``right'' ($y$) end of the link $xy$ and they obey the standard $SU(2)$ commutation relations at each end of the link
\begin{align}
\label{SO5LR}
&[L^a_{xy},L^b_{wz}] = 2 i \delta_{xw} \delta_{yz} \varepsilon_{abc} L^c_{xy}, \nonumber \\ 
&[R^a_{xy},R^b_{wz}] = 2 i \delta_{xw} \delta_{yz} \varepsilon_{abc} R^c_{xy}, \nonumber \\ 
&[L^a_{xy},R^b_{wz}] = 0.
\end{align}
In order to guarantee the correct gauge transformation properties of the quantum link operators (cf.\ Eq.\ (\ref{gaugeU})), we impose the commutation relations
\begin{align}
\label{SO5LRU}
&[L^a_{xy},U_{wz}] = - \delta_{xw} \delta_{yz} \sigma_a U_{xy}, \nonumber \\
&[R^a_{xy},U_{wz}] = \delta_{xw} \delta_{yz} U_{xy} \sigma_a.
\end{align} 
The same relations also hold in Wilson's lattice gauge theory, but they are realized in an infinite-dimensional link Hilbert space. This is unavoidable if one insists that $U_{xy}$ is an $SU(2)$ matrix with c-number valued matrix elements. As described above, the elements of a quantum link operator are non-commuting objects. In contrast to Wilson's theory, in order to realize exact $SU(2)$ gauge symmetry in a finite-dimensional link Hilbert space, we postulate the following non-trivial commutation relations
\begin{align}
\label{SO5UU}
&[U^0_{xy},U^0_{wz}] = 0, \nonumber \\ 
&[U^0_{xy},U^a_{wz}] = 2 i \delta_{xw} \delta_{yz} (R^a_{xy} - L^a_{xy}), \nonumber \\ 
&[U^a_{xy},U^b_{wz}] = 2 i \delta_{xw} \delta_{yz} \varepsilon_{abc} (R^c_{xy} + L^c_{xy}).
\end{align}
This closes the algebra of the four Hermitian quantum link operators $U^0_{xy}$, $U^a_{xy}$ and the six Hermitian operators $L^a_{xy}$ and $R^a_{xy}$, which turn out to be the generators of an embedding $SO(5)$ algebra. It is important to note that the commutation relations of Eq.\ (\ref{SO5UU}) do not compromise the gauge symmetry. In fact, lattice gauge theories with exact $SU(2)$ gauge invariance can now be constructed by choosing any representation of $SO(5)$ on each link. The embedding algebra $SO(5)$ contains $SO(4) = SU(2)_L \times SU(2)_R$ as a subalgebra, which gives rise to the gauge symmetry on each link. In particular, there is no $SO(5)$ but only an $SU(2)$ gauge symmetry.

States $|\Psi\rangle$ that belong to the physical Hilbert space must obey the Gauss law
\begin{align}
G^a_x |\Psi\rangle &= 0.
\end{align}
An external non-Abelian static charge, which can be characterized by an $SU(2)$ representation, violates the Gauss law at some lattice site $x$. If an external charge carries a half-integer representation of $SU(2)$, an unbreakable center $\Z(2)$ flux string emanates from it. Such a string can only end in another external charge also carrying a half-integer representation. External charges that carry integer representations, on the other hand, are not confined by an unbreakable string, because they can be screened by dynamical non-Abelian charges associated with the gauge field.

The above quantum link model construction naturally extends to $Sp(N)$ gauge theories with $N \geq 2$ \cite{BROWER2004149}. In that case, the embedding algebra is $Sp(2N)$. It should be noted that $SU(2) = Sp(1)$ and $Sp(2) = Spin(5)$ --- the universal covering group of $SO(5)$. Similarly, $SO(N)$ and $SU(N)$ quantum link models are realized with $SO(2N)$ and $SU(2N)$ embedding algebras, respectively \cite{BROWER2004149}.

\begin{figure}
	\begin{center}
		\includegraphics[width=0.23\textwidth]{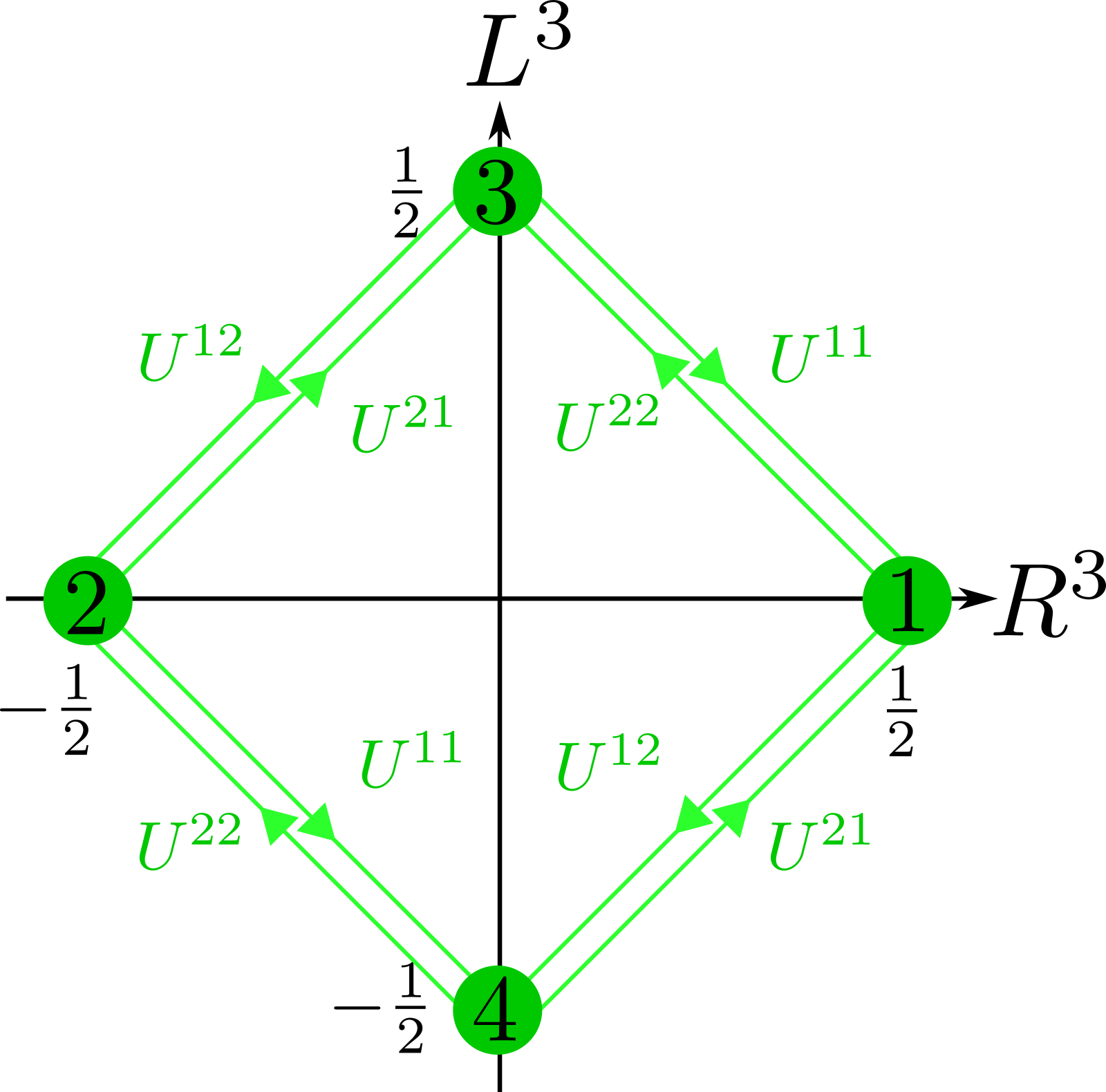}
		\hspace{0.01\textwidth} 
		\includegraphics[width=0.23\textwidth]{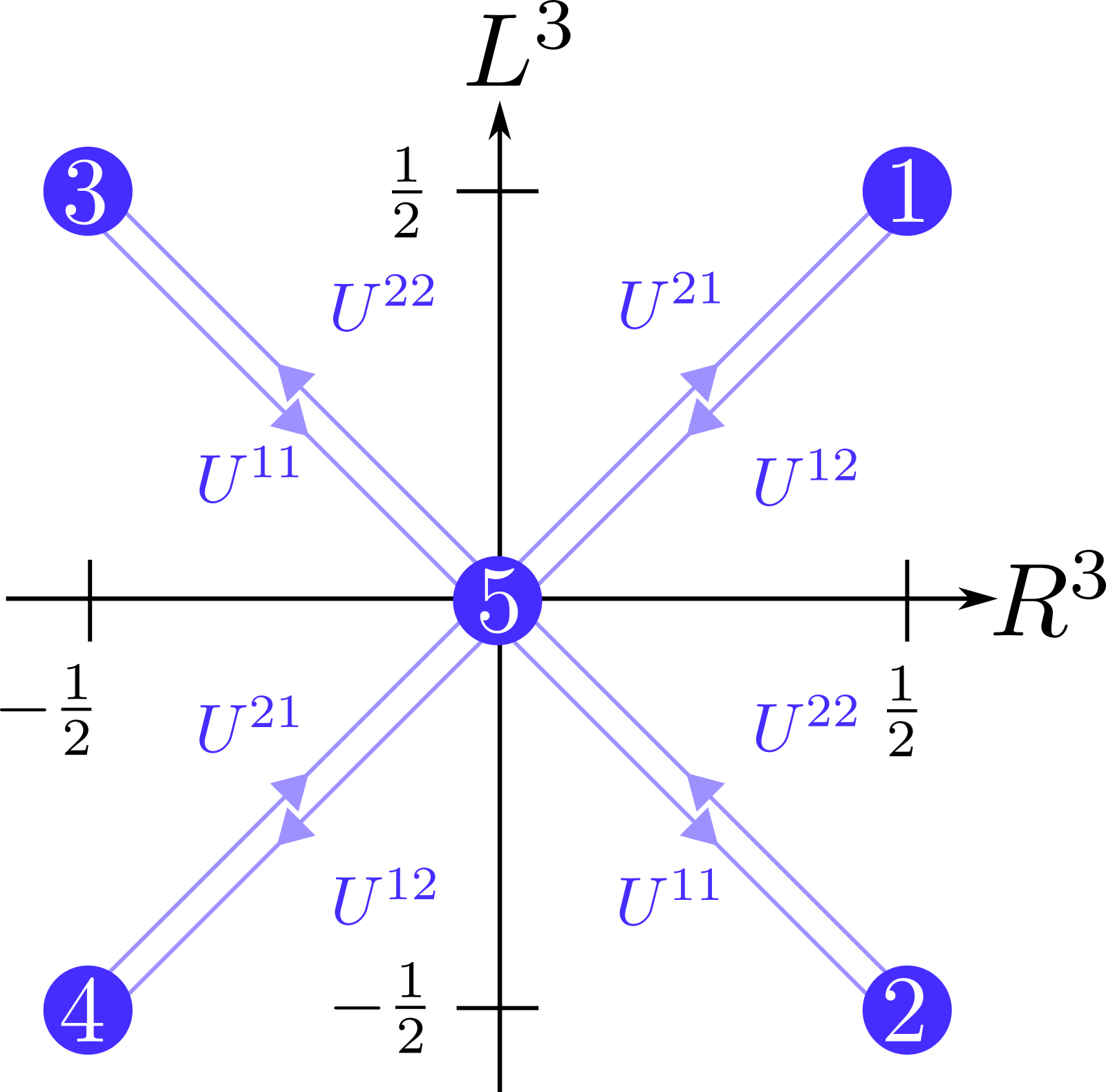}
	\end{center}
	\caption{\it Weight diagrams for the 4-dimensional spinor representation (left) and the 5-dimensional vector representation (right). The matrix elements $U^{ij}$ of the quantum link operator act as shift operators between the different states.}
	\label{Weightdiagrams}
\end{figure}

In the following, we will choose the smallest non-trivial representation of the embedding algebra $SO(5)$, namely the 4-dimensional spinor representation. In that case, the link Hilbert space is 4-dimensional. Alternatively, one could choose the 5-dimensional vector representation. The corresponding weight diagrams are illustrated in Fig.\,\ref{Weightdiagrams}. Under the $SU(2)_L \times SU(2)_R$ subgroup of $SO(5)$, the fundamental spinor and the vector representation decompose as
\begin{align}
&\{4\} = \{1,2\} + \{2,1\}, \nonumber \\ 
&\{5\} = \{1,1\} + \{2,2\}.
\end{align}
In particular, the vector (but not the spinor) representation carries the same $SU(2)$ representation, both on the left and on the right end of a link. The same feature is also inherent in Wilson's lattice gauge theory. In contrast to this, in the spinor representation one end of the link carries a singlet and the other end carries a doublet representation. As we will see, this feature, which is unique to quantum link models, gives rise to new non-Abelian confinement phases with crystalline order and fractionalized confining strings carrying delocalized $\Z(2)$ center electric flux.

Finally, let us construct a simple quantum link model Hamiltonian as
\begin{align}
H &= \frac{g^2}{2} \sum_{\langle xy \rangle} (L^a_{xy} L^a_{xy} + R^a_{xy} R^a_{xy}) + \frac{1}{2 g^2} \sum_{\hexagon} \mbox{Tr} U_{\hexagon},
\label{Hamiltonian}
\end{align}
where $U_{\hexagon}$ is the product of quantum link operators along the oriented boundary of a hexagon visiting the sites $x$, $y$, $z$, $u$, $v$, $w$ in cyclic order, i.e.
\begin{align}
U_{\hexagon} &= U_{xy} U_{yz} U_{zu} U_{uv} U_{vw} U_{wx}.
\end{align}
The term proportional to $g^2$ represents the electric field energy, and the hexagon-plaquette term represents the magnetic field energy. This form of the Hamiltonian is exactly the same as in Wilson's lattice gauge theory, except that the link and electric field operators are represented differently. It is straightforward to convince oneself that $H$ is indeed gauge invariant, i.e.\ $[H,G^a_x] = 0$, for the local generators $G^a_x$ of gauge transformations at all sites $x$ (cf.\ Eq.\ (\ref{gaugetrafo})). It should be noted that for the spinor representation $\{4\}$ the electric field energy is a trivial constant, while for the vector representation $\{5\}$ it distinguishes the non-zero electric flux states 1, 2, 3, 4 from the zero flux state 5 (cf.\ Fig.\,\ref{Weightdiagrams}).

\subsection{Rishon Representation of the $SU(2)$ Quantum Link Model}

In contrast to Wilson's lattice gauge theory, quantum link models allow a factorization of the quantum link operators into so-called rishon constituents.  For the $SU(2)$ quantum link model these are color-doublet fermions residing at the ends of a link that obey standard anti-commutation relations
\begin{align}
&\{{c^i}_{xy,+}^\dagger,c^j_{wz,+}\} = \{{c^i}_{xy,-}^\dagger,c^j_{wz,-}\} =\delta_{xw} \delta_{yz} \delta_{ij}, \nonumber\\ 
&\{{c^i}_{xy,+}^\dagger,c^j_{wz,-}\} = \{{c^i}_{xy,-}^\dagger,c^j_{wz,+}\} = 0, \nonumber \\ 
&\{{c^i}_{xy,\pm}^\dagger,{c^j}_{wz,\pm}^\dagger\} = \{c^i_{xy,\pm},c^j_{wz,\pm}\} = 0.
\end{align}
Here $i$ and $j$ are $SU(2)$ color indices, and $-$ and $+$ refer to the $x$ and $y$ ends of the link $\langle xy \rangle$, respectively. In the rishon representation, the electric flux operators residing on a link are given by
\begin{align}
&L^a_{xy} = \frac{1}{2} {c^i}_{xy,-}^\dagger \sigma^a_{ij} c^j_{xy,-}, \nonumber\\ 
&R^a_{xy} = \frac{1}{2} {c^i}_{xy,+}^\dagger \sigma^a_{ij} c^j_{xy,+},
\end{align}
and the matrix elements $U_{xy}^{ij}$ of a quantum link matrix $U_{xy}$ take the form
\begin{align}
&U_{xy}^{11}={c^1}_{xy,+}^\dagger c^1_{xy,-} + {c^2}_{xy,-}^\dagger c^2_{xy,+}, \nonumber \\
&U_{xy}^{12}= {c^2}_{xy,+}^\dagger c^1_{xy,-} - {c^2}_{xy,-}^\dagger c^1_{xy,+}, \nonumber \\ 
&U_{xy}^{21}={c^1}_{xy,+}^\dagger c^2_{xy,-} - {c^1}_{xy,-}^\dagger c^2_{xy,+}, \nonumber \\ 
&U_{xy}^{22}={c^2}_{xy,+}^\dagger c^2_{xy,-} + {c^1}_{xy,-}^\dagger c^1_{xy,+}.
\end{align}
It is straightforward to show that they indeed satisfy the $SO(5)$ commutation relations of Eqs.\ \eqref{SO5LR}-\eqref{SO5UU}. One sees that the quantum link operator $U_{xy}$ shuffles a rishon from one end of the link to the other, keeping the total number of rishons per link,
\begin{align}
{\cal N}_{xy} &= {c^i}_{xy,+}^\dagger c^i_{xy,+} + {c^i}_{xy,-}^\dagger c^i_{xy,-},
\end{align}
fixed.

The spinor representation $\{4\} = \{1,2\} + \{2,1\}$ has ${\cal N}_{xy} = 1$ rishon per link, which resides either on its left or on its right end.  The vector representation $\{5\} = \{1,1\} + \{2,2\}$, on the other hand, has ${\cal N}_{xy} = 2$ rishons per link, which reside on opposite ends of the link for the states 1, 2, 3, 4 in Fig.\,\ref{Weightdiagrams}.  When the two rishons sit on the same end of the link, due to their fermionic nature, they necessarily form a color-singlet.  The symmetric superposition of a two-rishon singlet sitting on the left and on the right end of the link corresponds to the state 5 in Fig.\,\ref{Weightdiagrams}.  The anti-symmetric superposition, on the other hand, is an $SO(5)$ singlet and thus decouples from the quantum link model dynamics.  

\begin{figure}
	\begin{center}
		\includegraphics[width=0.45\textwidth]{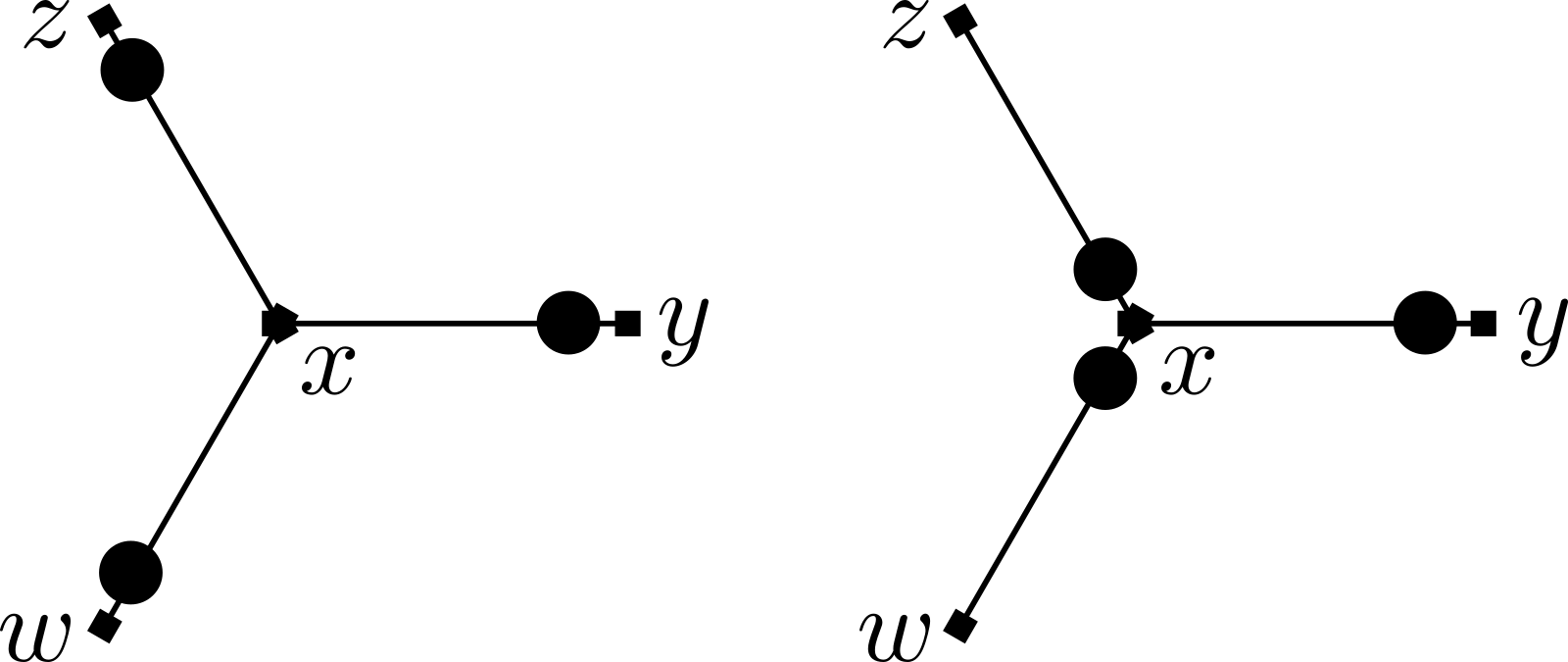}
	\end{center}
	\caption{\it The Gauss law at a site $x$ is satisfied in the $\{4\}$-representation if either no rishons (left) or two rishons (right) reside next to the site $x$.}
	\label{Gausslaw}
\end{figure}

\begin{figure}
	\begin{center}
		$\underset{\{4\}-\rm{Representation}}{\includegraphics[width=0.2\textwidth]{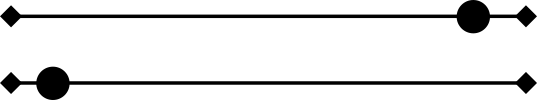}}$
		\hspace{0.05\textwidth} 
		$\underset{\{5\}-\rm{Representation}}{\includegraphics[width=0.2\textwidth]{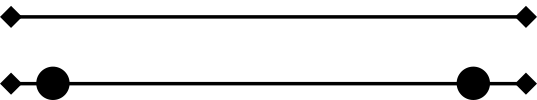}}$
	\end{center}
	\caption{\it Possible configurations for the color $1/2$ rishons on a link for the $\left\{4\right\}$-representation (left) and for the $\left\{5\right\}$-representation (right).}
	\label{FIG:RishonLinkRep}
\end{figure}

Let us now discuss the realization of the Gauss law in the rishon representation. Since on the honeycomb lattice three links emanate from a site, in the $\{4\}$-representation up to three rishons may reside next to a lattice site. The Gauss law requires that they form a local color-singlet. Since every rishon represents a color-doublet, only zero or two (but not one or three) rishons can meet at a site. This is illustrated in Fig.\,\ref{Gausslaw}. The explicit realization of the Gauss law allows us to work in a manifestly gauge invariant basis of physical states, which have an even number of rishons next to each lattice site. In this basis, the color-state of the rishons is implicitly determined because each rishon-pair at a vertex must form a color-singlet. This implies that the dimension of the local link Hilbert space is effectively reduced from 4 to 2.  In particular, in Fig.\,\ref{Gausslaw} it is sufficient to specify whether a rishon resides on the left or right end of a link. Its color state is determined by the fact that it forms a color-singlet with its rishon partner next to the same site. The reduced link states for the $\{4\}$-representation are illustrated on the left-hand side of Fig.\,\ref{FIG:RishonLinkRep}.

A similar situation arises for the $\{5\}$-representation. The states 1, 2, 3, 4 have one rishon at each end of a link, which must then form a color-singlet with another rishon residing on an adjacent link that is also in one of the states 1, 2, 3, 4. In the state 5 both rishons sit on the same end of a link and form a color-singlet by themselves. Hence, such a link, which does not carry electric flux, does not contribute to the Gauss law. The reduced link states of the $\{5\}$-representation are illustrated on the right-hand side of Fig.\,\ref{FIG:RishonLinkRep}. State 5 carries a color-singlet at both ends of the link and is represented by an empty link.

\subsection{Non-Abelian External Charges, $\Z(2)$ Center Symmetry, and Electric Flux Strings}
\label{SUBSEC:ElecFlux}

As in any gauge theory, one can introduce external charges by violating Gauss' law at a specific position. In $SU(2)$ gauge theory, the external charges are characterized by their $SU(2)$ representation. In particular, there are charges either with an integer or with a half-integer color-spin representation.  Pairs of non-Abelian external charges that carry a half-integer representation are connected by an unbreakable $\Z(2)$ center electric flux string. The string that connects charges with an integer representation, on the other hand, can break by the pair creation of dynamical charges. Since at most three color-doublets can sit near a site, when one uses the $\{4\}$- or $\{5\}$-representation of the $SO(5)$ embedding algebra one is limited to external charges $1/2$, $1$, or $3/2$.

\begin{figure}
  \begin{center}
    \includegraphics[width=0.45\textwidth]{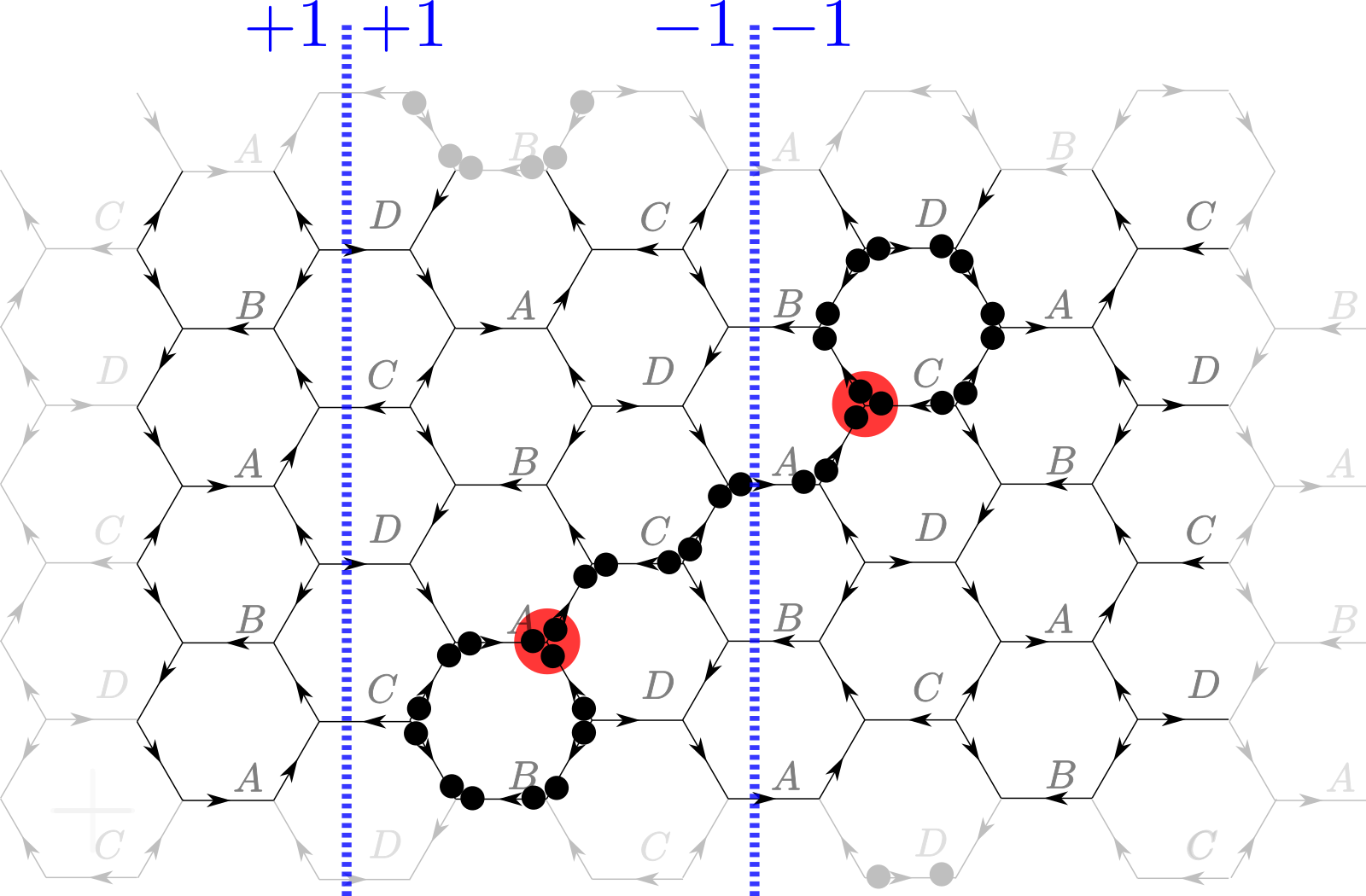}\\
    \vspace{10pt} 
    \includegraphics[width=0.45\textwidth]{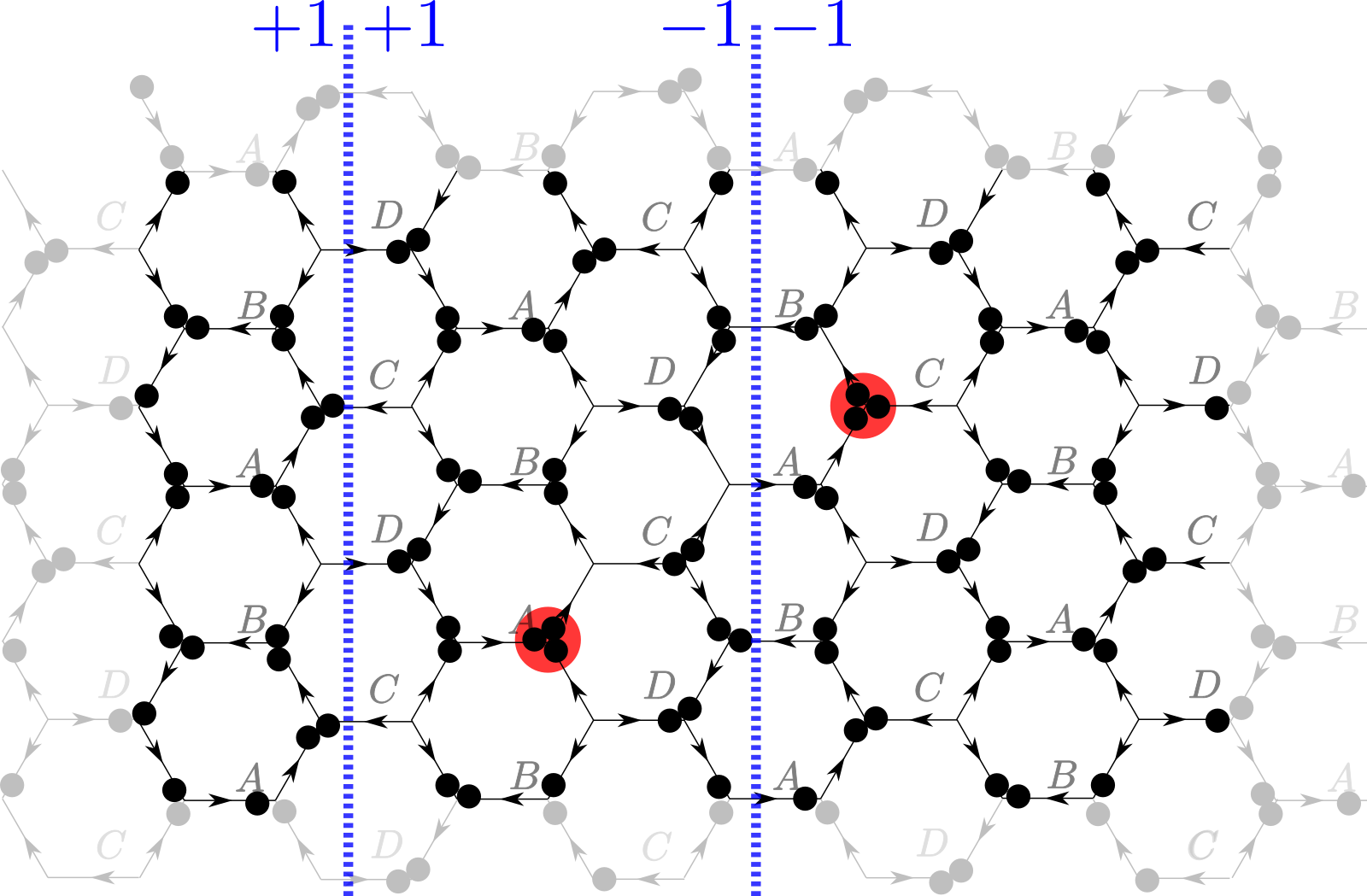}
  \end{center}
  \caption{\it Pairs of external non-Abelian static charges carrying a $3/2$ representation of $SU(2)$ (red filled circles) are included in the $\{5\}$-representation (top) and in the $\{4\}$-representation (bottom) on a periodic lattice. Electric flux is identified via the closed loops (vertical dotted lines) by a change in the center symmetry values from $+1|+1$ (no flux) to $-1|-1$ (flux).}
  \label{Flux}
\end{figure}

When one uses the $\{5\}$-representation, in the absence of external charges the Gauss law implies that flux-carrying links in the states 1, 2, 3, 4 form closed loops. A pair of external non-Abelian static charges carrying a half-integer representation of $SU(2)$ is thus connected by an unbreakable $\Z(2)$ flux string of states 1, 2, 3, 4. This is illustrated in Fig.\,\ref{Flux} (top). A similar situation arises in Wilson's standard lattice gauge theory \cite{PhysRevLett.102.191601}. In Fig.\,\ref{Flux} (top) the closed loops (vertical dotted lines) wrapping around the periodic volume on the dual lattice are used to measure the $\Z(2)$ flux that goes through them. For this purpose, one considers all links that cross the corresponding loop. If the total number of rishons on one side of the loop is even, no net $\Z(2)$ flux goes through the loop. In Fig.\,\ref{Flux} (top) this is the case for the loop on the left. The numbers $+1|+1$ indicate that the rishon-count on both sides of the loop is even. If the number of rishons is odd, on the other hand, one unit of $\Z(2)$ flux crosses the loop.  This is the case for the loop on the right, for which $-1|-1$ indicates an odd-odd rishon-count. This definition of the flux is consistent, because the total number of rishons is the same on both sides of the loop. It should be noted that the closed loop can be deformed arbitrarily (on the dual lattice) without changing the result, as long as no external half-integer charges are crossed. As we will discuss in the next subsection, the Hamiltonian respects the $\Z(2)$ center symmetry.

Interestingly, a rather different situation arises when we formulate the $SU(2)$ quantum link model using the $\{4\} = \{1,2\} + \{2,1\}$ representation of the embedding algebra $SO(5)$. Because it has only one rishon per link, unlike the $\{5\} = \{1,1\} + \{2,2\}$ representation, the $\{4\}$-representation does not carry the same $SU(2)$ representation on the left and on the right. As a consequence, $\Z(2)$ center electric flux is then not uniquely localizable on a given link.  Still, the concept of unbreakable $\Z(2)$ flux strings connecting half-integer external charges applies here as well. However, the $\Z(2)$ flux is delocalized.  This is illustrated in Fig.\,\ref{Flux} (bottom). The flux is defined as before. However, in case of the $\{4\}$-representation it is important that the lattice has an even extent. Otherwise, the number of rishons modulo 2 on the left- and on the right-hand side of the closed loop would be different.

Although in the $\{4\}$-representation $\Z(2)$ flux is no longer localized, one can determine the total $\Z(2)$ flux flowing through a closed loop that wraps around the periodic volume. In Fig.\,\ref{Flux} (bottom) there are again two loops (vertical dotted lines) that are closed over the periodic spatial boundary on the dual lattice. Net $\Z(2)$ flux flows only through the loop on the right, indicating that $\Z(2)$ flux indeed connects the two external charges, despite the fact that (in contrast to the $\{5\}$-representation case) one cannot tell which links actually carry the flux. Although the flux itself is delocalized, from our Monte Carlo results we will conclude that its energy is carried by two fractionalized strands, which play the role of domain walls separating distinct crystalline confined phases. This situation is unique to quantum link models and does not arise in Wilson's formulation of lattice gauge theory.

\subsection{Classification of Ring-Exchange Hamiltonians}

Although the Hamiltonian of Eq.\ (\ref{Hamiltonian}) is a natural choice, it is not the most general one. In this subsection, we construct the most general $SU(2)$ gauge invariant ring-exchange hexagon-plaquette Hamiltonian that respects the lattice symmetries. Here we restrict ourselves to the spinor representation $\{4\}$ of the $SO(5)$ embedding algebra.

\begin{figure}
	\begin{center}
		$\underset{\textit{Environment $E=1$}}{\includegraphics[width=0.22\textwidth]{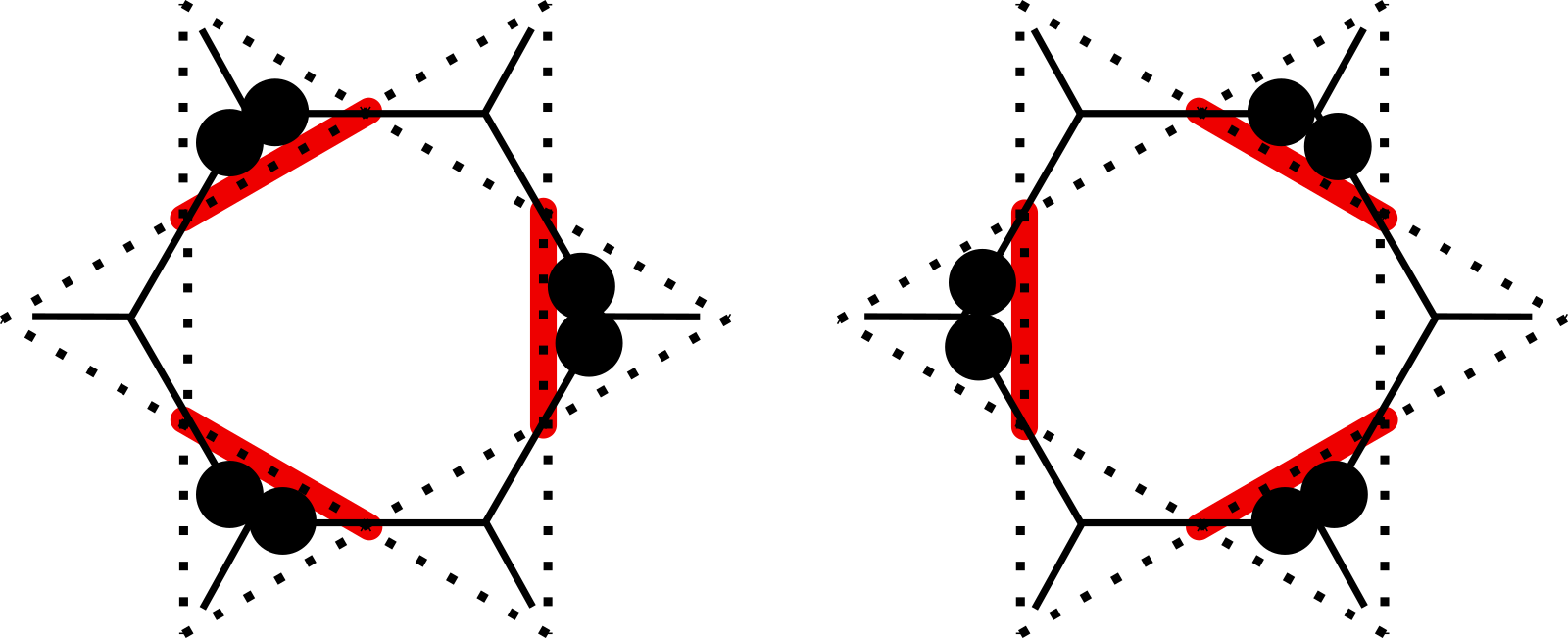}}$
		\hspace{0.025\textwidth}
		$\underset{\textit{Environment $E=2$}}{\includegraphics[width=0.22\textwidth]{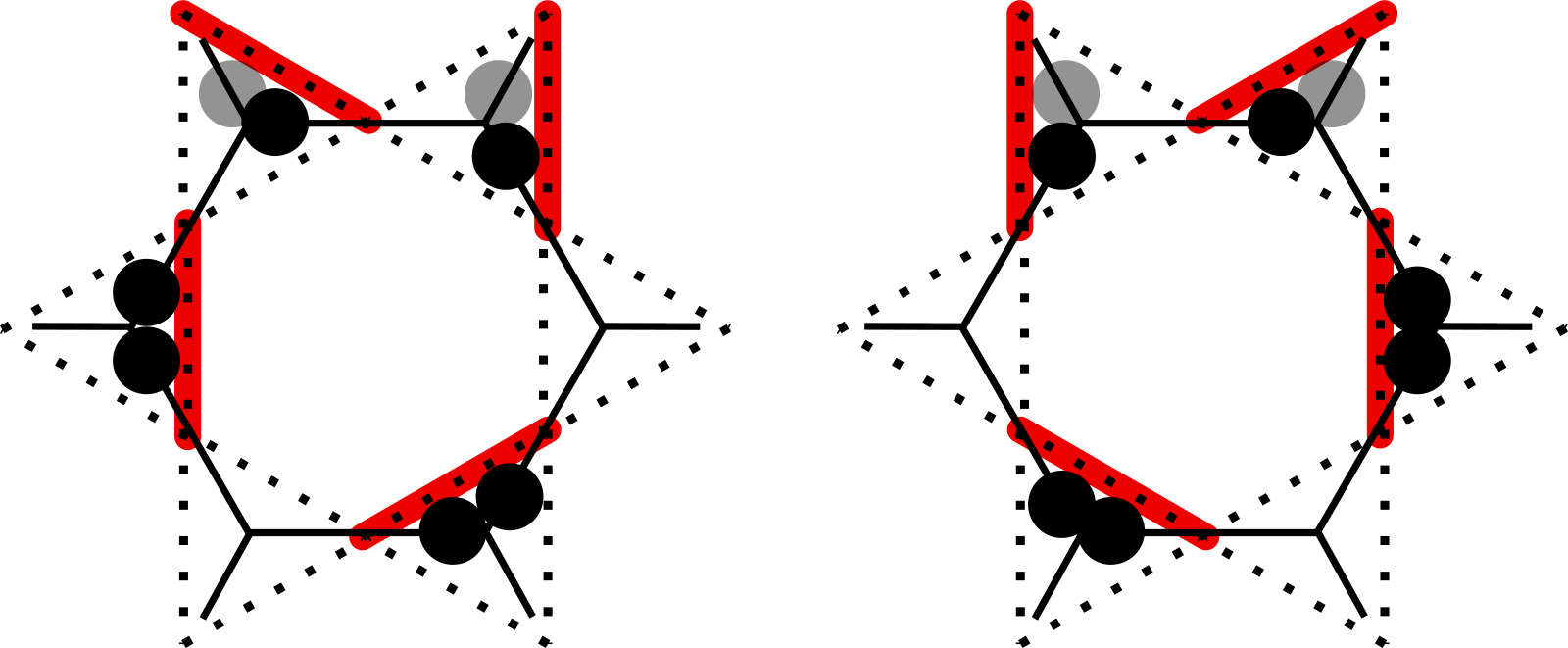}}$\\
		\vspace{15pt}
		$\underset{\textit{Environment $E=3$}}{\includegraphics[width=0.22\textwidth]{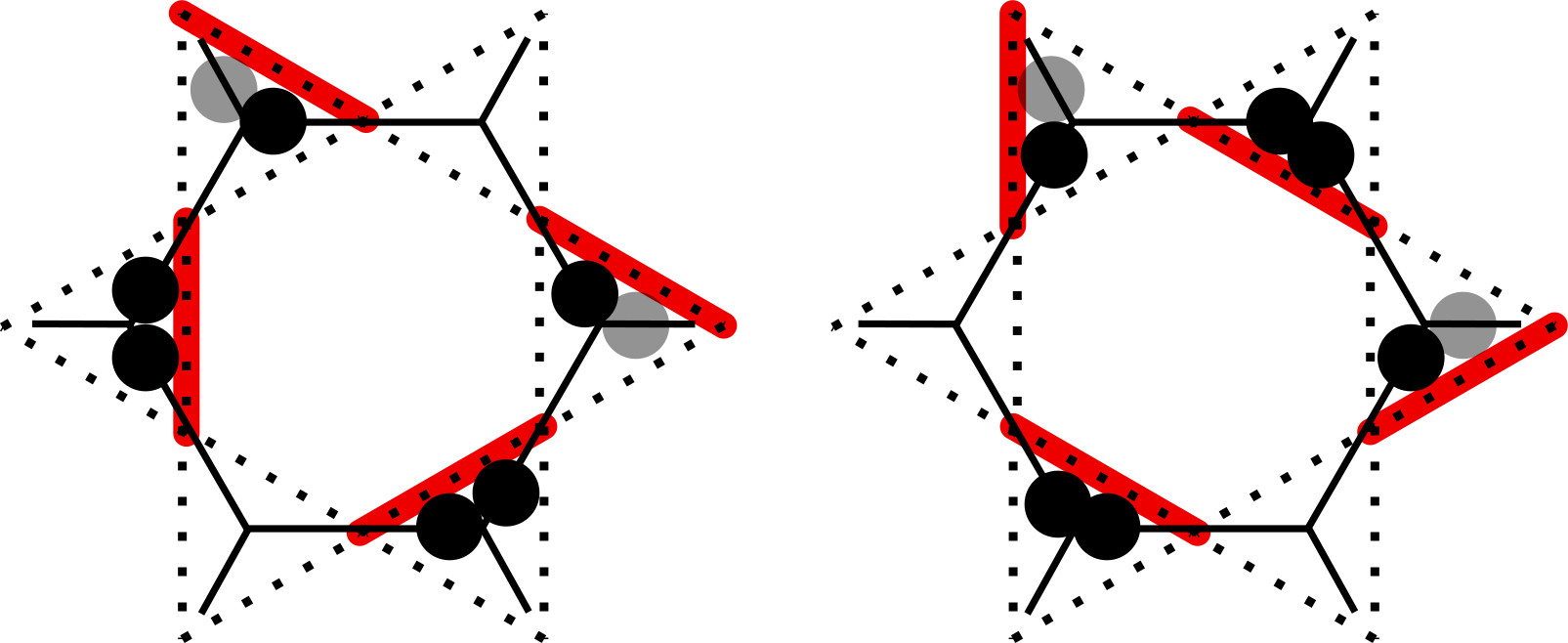}}$
		\hspace{0.025\textwidth}
		$\underset{\textit{Environment $E=4$}}{\includegraphics[width=0.22\textwidth]{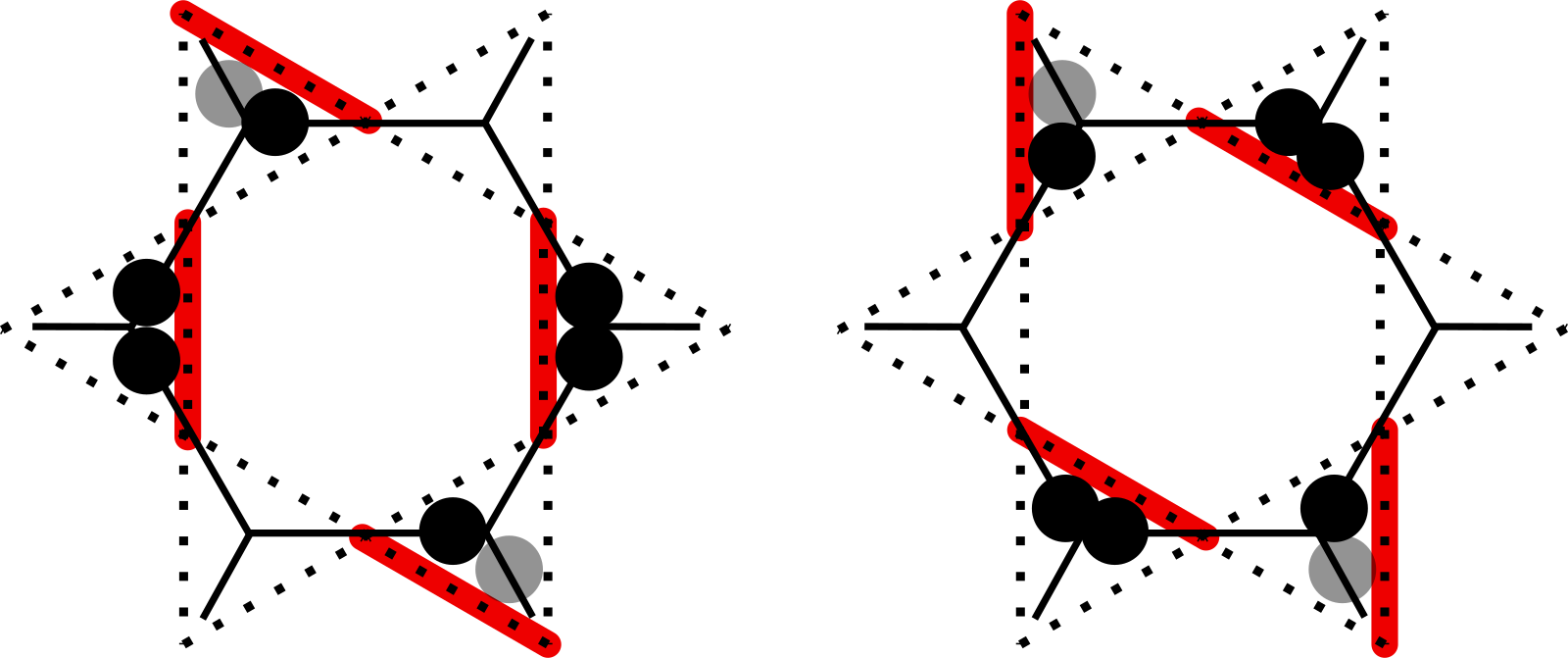}}$\\
		\vspace{15pt}
		$\underset{\textit{Environment $E=5$}}{\includegraphics[width=0.22\textwidth]{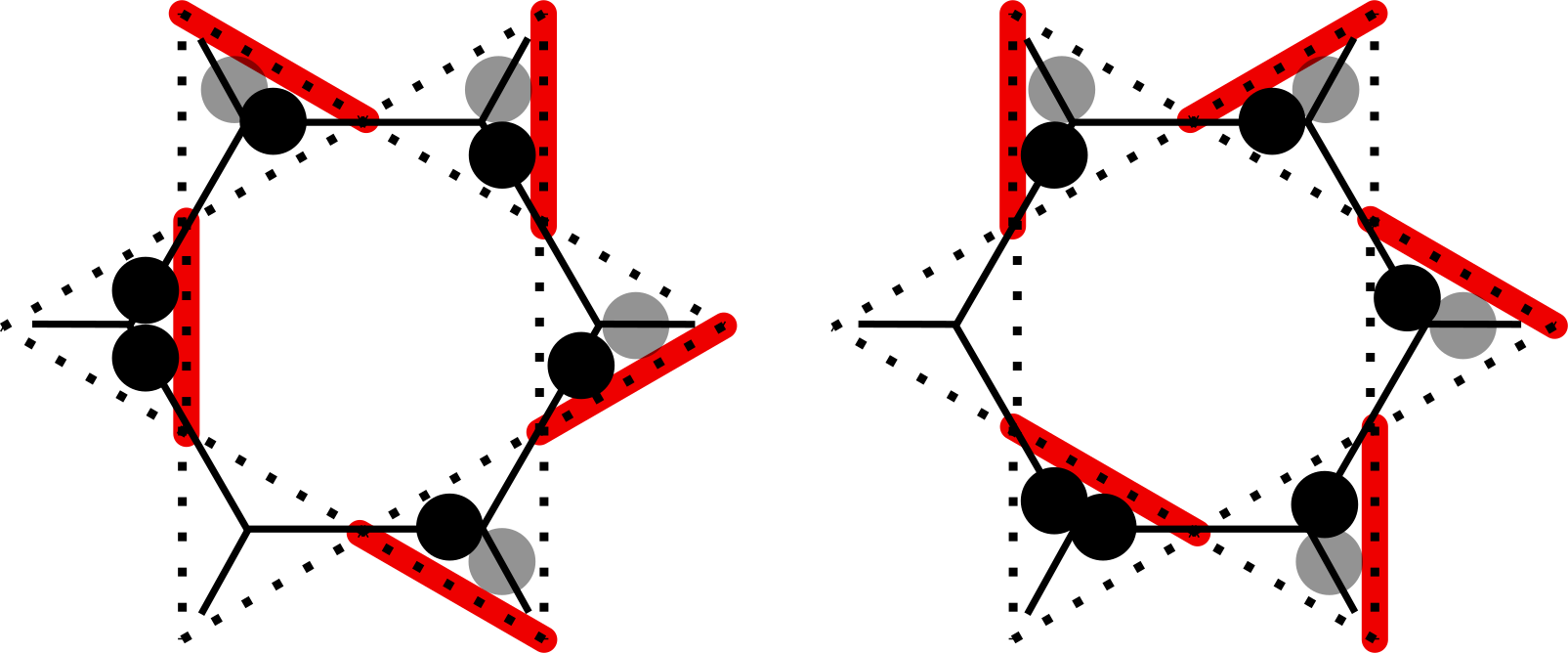}}$
		\hspace{0.025\textwidth}
		$\underset{\textit{Environment $E=6$}}{\includegraphics[width=0.22\textwidth]{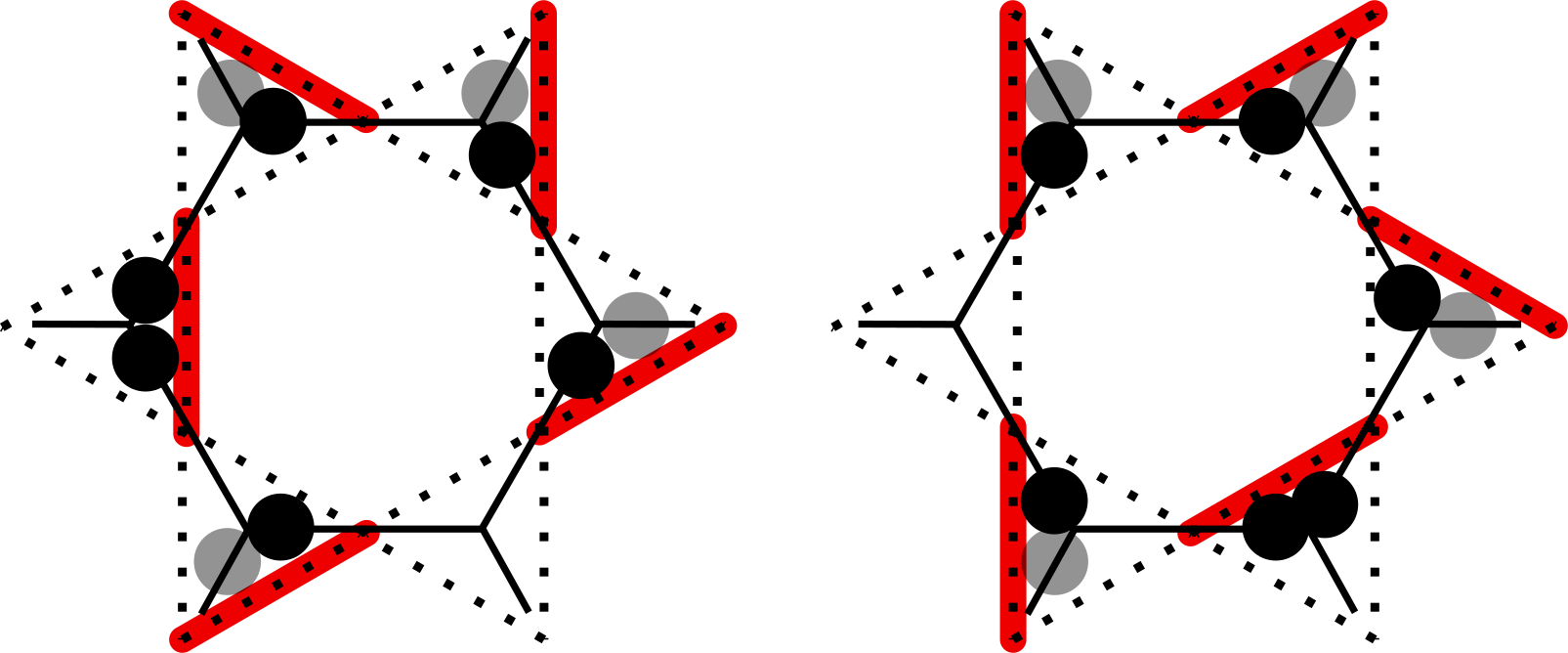}}$\\
		\vspace{15pt}
		$\underset{\textit{Environment $E=7$}}{\includegraphics[width=0.22\textwidth]{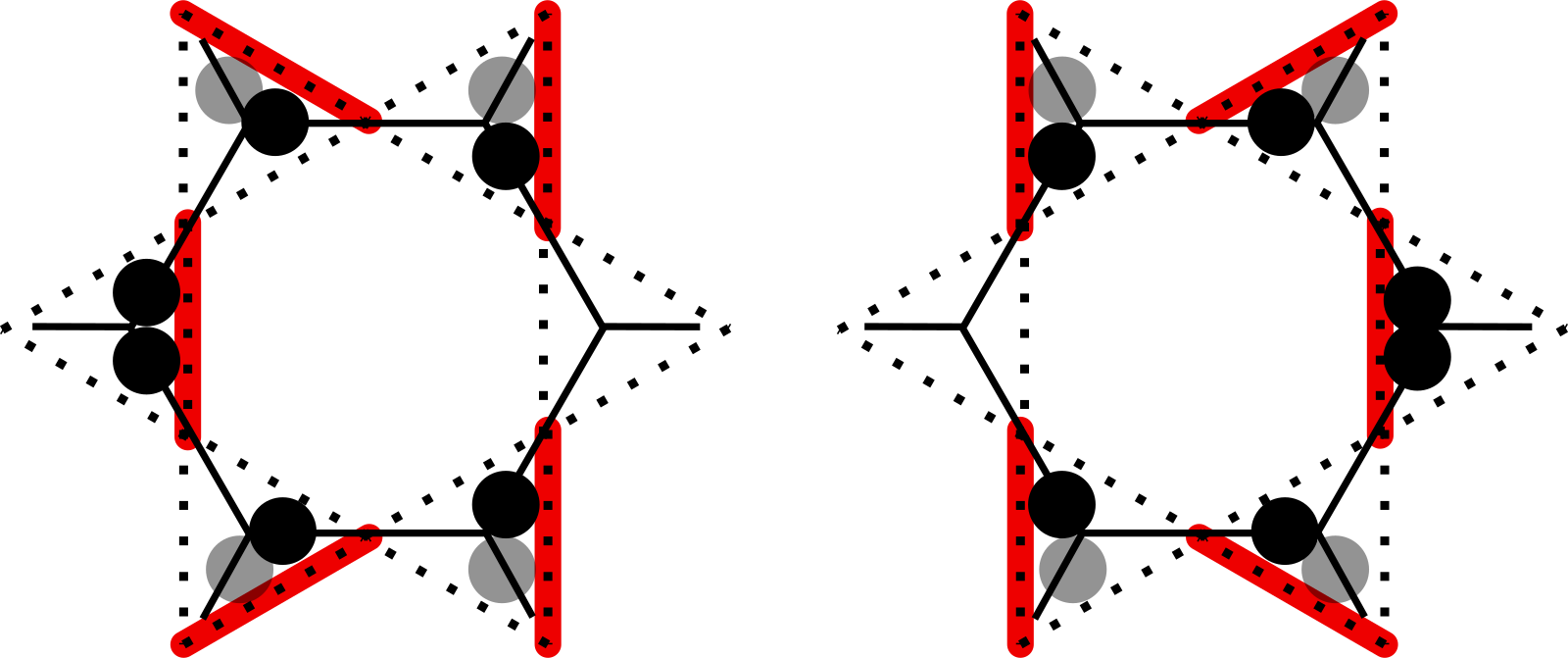}}$
		\hspace{0.025\textwidth}
		$\underset{\textit{Environment $E=8$}}{\includegraphics[width=0.22\textwidth]{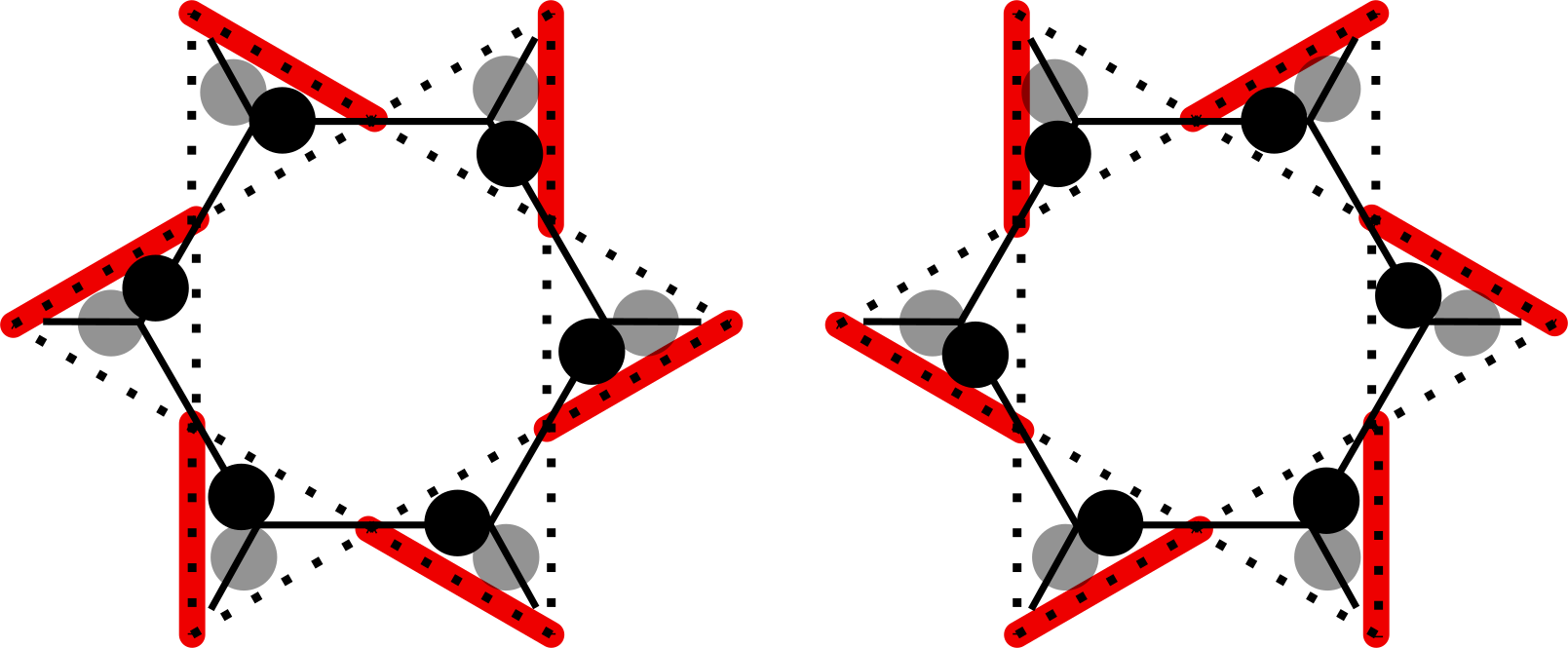}}$
	\end{center}
	\caption{\it The eight distinct environments (semi-transparent rishons) together with their allowed rishon configurations (black rishons) for the spinor representation $\{4\}$.  The corresponding dimer configurations (bold red bars) are shown on the Kagom\'e lattice (dotted links). The environment $E=8$ defines two pinwheel plaquettes with opposite pinwheel orientations.}
	\label{Ringexchange}
\end{figure}

Let us consider a single hexagon with six internal links connecting the vertices and with six external links attached to these vertices from outside. The positions of the rishons on the external links define an environment in which the hexagon-plaquette is embedded. In particular, in order to satisfy the Gauss law, the internal rishons must be positioned in such a way that the total number of rishons at each lattice site is even. Up to lattice rotations and reflections, this defines eight distinct cases which are illustrated in Fig.\,\ref{Ringexchange}. Each of the eight environments allows two rishon configurations which are related by a ring-exchange process that moves all six internal rishons from one end of their link to the other.

Since the gauge invariant state of each link is specified by the position of the rishon, either at the left or at the right end of a link, the hexagon-plaquette Hamiltonian is a $2^6 \times 2^6 = 64 \times 64$ matrix, which decomposes into 32 blocks of $2 \times 2$ matrices. Each $2 \times 2$ matrix
\begin{align}
H_E &= \left(\begin{array}{cc} W^1_E & T_E \\ T_E^* & W^2_E \end{array} \right)
\end{align}
is Hermitian and corresponds to a ring-exchange transition in a particular environment that is characterized by the type of ring-exchange $E \in \{1,2,\dots,8\}$ (cf.\ Fig.\,\ref{Ringexchange}). The four real parameters contained in $W^a_E \in \R$ and $T_E \in \C$ are further restricted by lattice symmetries. These include 60 degree rotations around the center of a hexagon and reflections on axes going through that center. For all environments except $E = 3$, this implies that $W^1_E = W^2_E$ and $T_E \in \R$. For $E = 3$, no such restriction arises. Hence, the most general gauge-, rotation-, and reflection-invariant Hamiltonian has $7 \times 2 + 4 = 18$ real parameters (of which one overall additive constant is trivial).

It is now easy to see that the Hamiltonian respects the $\Z(2)$ center symmetry that we discussed in the previous subsection. The ring-exchange processes shift all rishons that reside on the six links of a hexagon from one end of a link to the other. Since the dual dotted lines in Fig.\,\ref{Flux}, which are used to identify the $\Z(2)$ flux, necessarily cross two links of a given hexagon, the number of rishons on both sides of the dotted line remains the same modulo 2.

\subsection{Relation to the Quantum Dimer Model on the Kagom\'e Lattice}

\begin{figure}
	\begin{center}
		\includegraphics[width=0.30\textwidth]{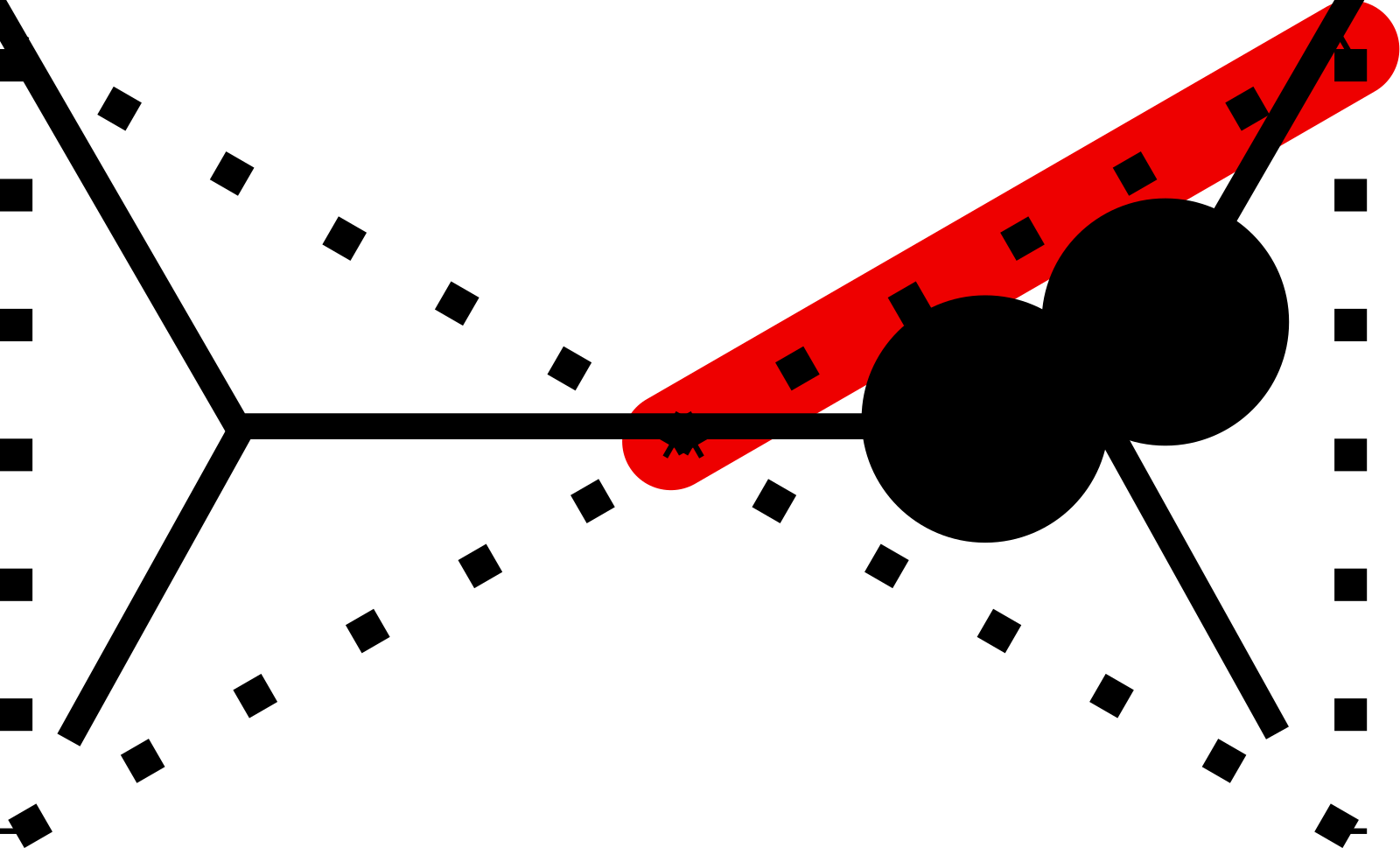}
	\end{center}
	\caption[Kagom\'e dimer and honeycomb rishons.]{\it Two rishons on the honeycomb lattice (solid links), forming a singlet pair, are identified with a corresponding dimer (red bold line) on the Kagom\'e lattice (dotted links).}
	\label{FIG:KagomeDimer}
\end{figure}

Interestingly, in the absence of external charges, the $SU(2)$ quantum link model on the honeycomb lattice, with the $\{4\}$-representation on each link, is equivalent to the quantum dimer model on the Kagom\'e lattice. As illustrated in Fig.\,\ref{FIG:KagomeDimer}, two rishons residing on neighboring links, which form a singlet in order to satisfy the Gauss law at a site, are identified with a dimer.
These dimers are naturally associated with the bonds that connect the centers of neighboring links. These bonds then form a Kagom\'e lattice. Since in the $\{4\}$-representation each link carries one rishon, there is exactly one dimer that touches a link center, which corresponds to a site of the Kagom\'e lattice.  Hence, a dimer configuration on the Kagom\'e lattice (resulting from the construction described above) automatically satisfies a dimer covering constraint.

As illustrated in Fig.\,\ref{Ringexchange}, the eight types of ring-ex\-chan\-ges, that contribute to the Hamiltonian of the $SU(2)$ quantum link model on the honeycomb lattice, are equivalent to the eight dimer moves that are usually considered on the Kagom\'e lattice \cite{PhysRevB.51.8318,PhysRevB.67.214413,PhysRevB.81.214413,PhysRevB.81.180402,PhysRevB.90.195148,PhysRevB.90.100406}. These are sometimes referred to as resonance moves. In particular, the environment $E = 8$ gives rise to a dimer pinwheel with two possible orientations. 

It should be pointed out that based on the arrow representation on the honeycomb lattice, which is designed specifically to characterize 
quantitatively the quantum dimer model on the 
Kagom\'e lattice \cite{PhysRevB.48.13647}, a $\Z_2$ lattice gauge theory was constructed in \cite{PhysRevB.87.104408}. On the one hand, it 
is demonstrated here that without the presence of external charges, the $SU(2)$ quantum link model is equivalent to the quantum dimer 
model on the Kagom\'e lattice. On the other hand, a $\Z_2$ lattice gauge theory was built on exactly the same quantum dimer model on 
the Kagom\'e lattice. It is interesting that such a correspondence exists between the $SU(2)$ quantum link model and a $\Z_2$ 
lattice gauge theory. This connection has not been appreciated previously in the literature. 

\subsection{Pinwheel Phases}

\begin{figure}
	\begin{center}
		\includegraphics[width=0.45\textwidth]{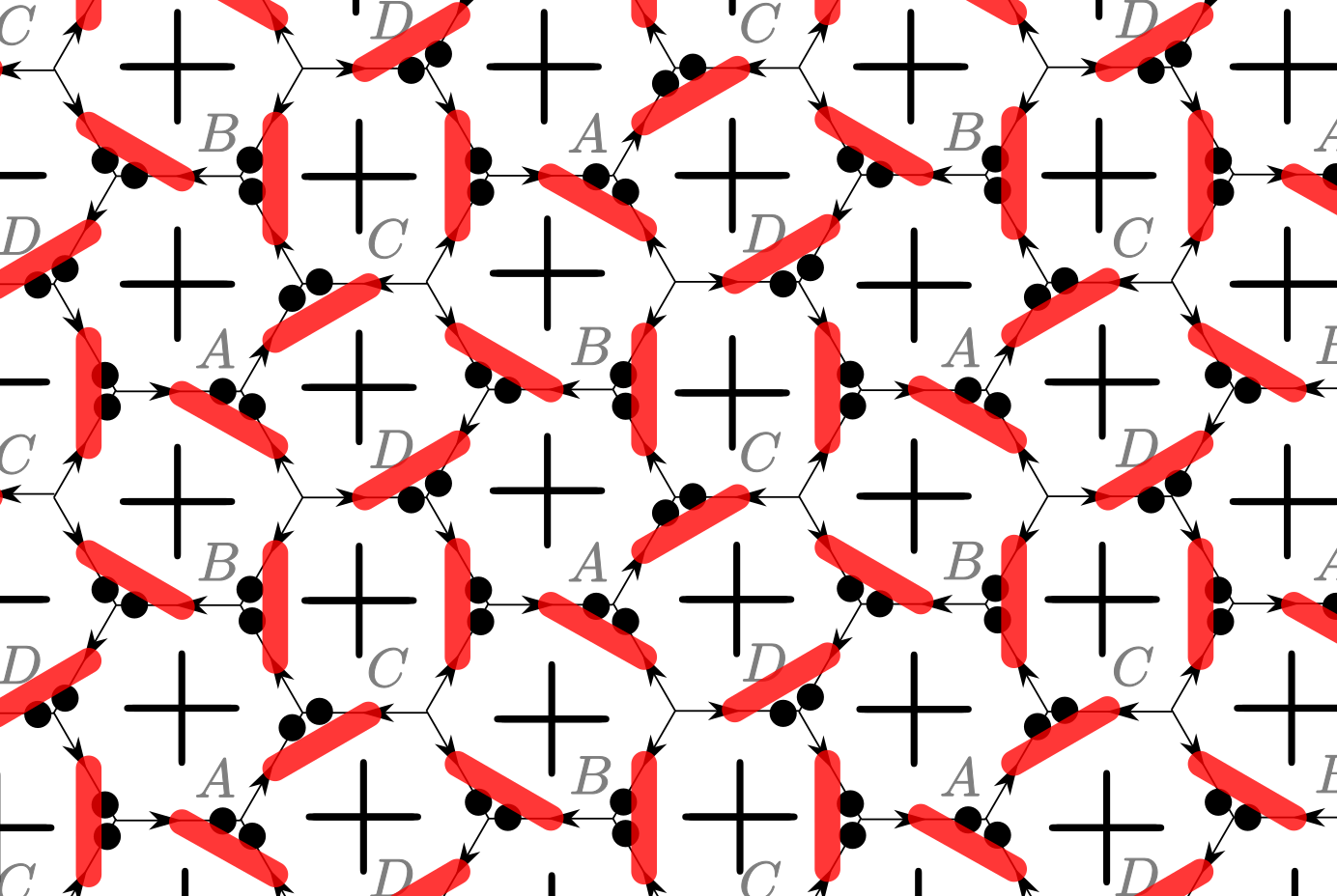}\\				\hspace{0.05\textheight}
		\includegraphics[width=0.45\textwidth]{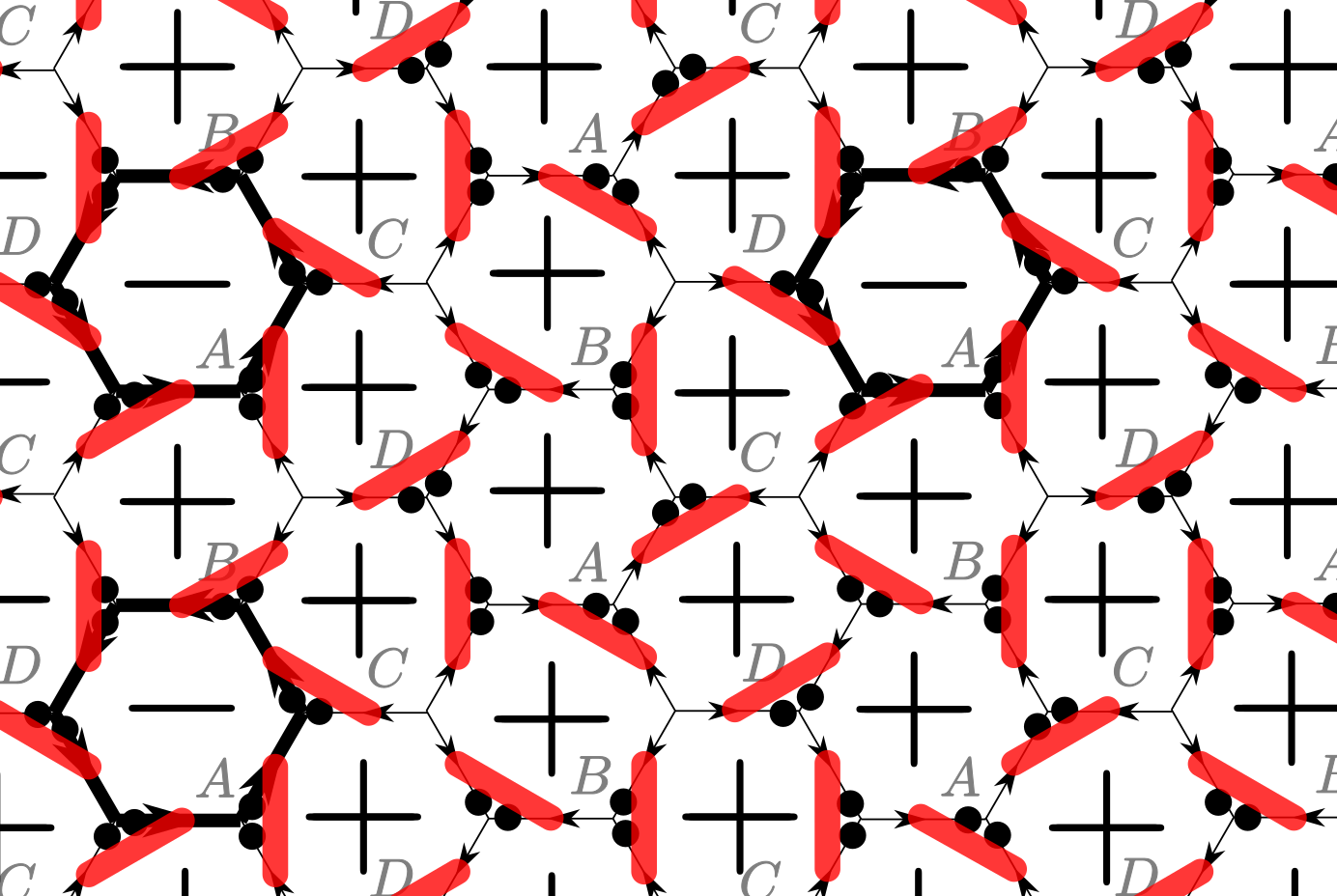}
	\end{center}
	\caption[Correlated and uncorrelated pinwheel phase]{\it ``Cartoon'' states of the correlated (top) and uncorrelated (bottom) pinwheel phase residing on sublattice $A$. The corresponding height variable configurations (defined in Subsection \ref{SUB:Height}) are denoted by $\pm$ signs associated with the hexagon centers. For the correlated pinwheel phase, all pinwheels are oriented in the same direction and all height variables are +. For the uncorrelated pinwheel phase, pinwheel plaquettes marked with a bold boundary carry a negative height variable and are oriented in the opposite direction.}
	\label{FIG:PinwheelPhases}
\end{figure}

We will not explore the entire 18-dimensional parameter space of the quantum link model. Instead, we focus on investigating the phase diagram as a function of $\lambda = T_4/T_8$, while setting all other $T_E = 0$. In addition, we put $W^a_E = 0$ for all environments $E$. These restrictions put us in phases with pinwheel order, which are illustrated in Fig.\,\ref{FIG:PinwheelPhases}.
The pinwheels reside on one of the four sublattices $A$, $B$, $C$, or $D$. As we will see later, for $T_4/T_8 > \lambda_c = 0.7185(5)$ there is a correlated pinwheel phase in which all pinwheels are oriented in the same direction. For $T_4/T_8 < \lambda_c$, on the other hand, pinwheels still persist on a given sublattice, but their orientations are no longer correlated. As we will demonstrate, the phase transition that separates the correlated from the uncorrelated pinwheel phase is second order in the universality class of the 3-d Ising model.

Since pinwheels exist on one of the four sublattices, both pinwheel phases spontaneously break lattice translation invariance and thus represent non-Abelian crystalline confined phases. In addition, the correlated pinwheel phase also breaks reflection symmetry. This $\Z(2)$ symmetry is restored in the uncorrelated pinwheel phase. Hence, it is not surprising that the corresponding phase transition is in the 3-d Ising universality class.  

\begin{figure}
  \begin{center}
    \includegraphics[width=0.45\textwidth]{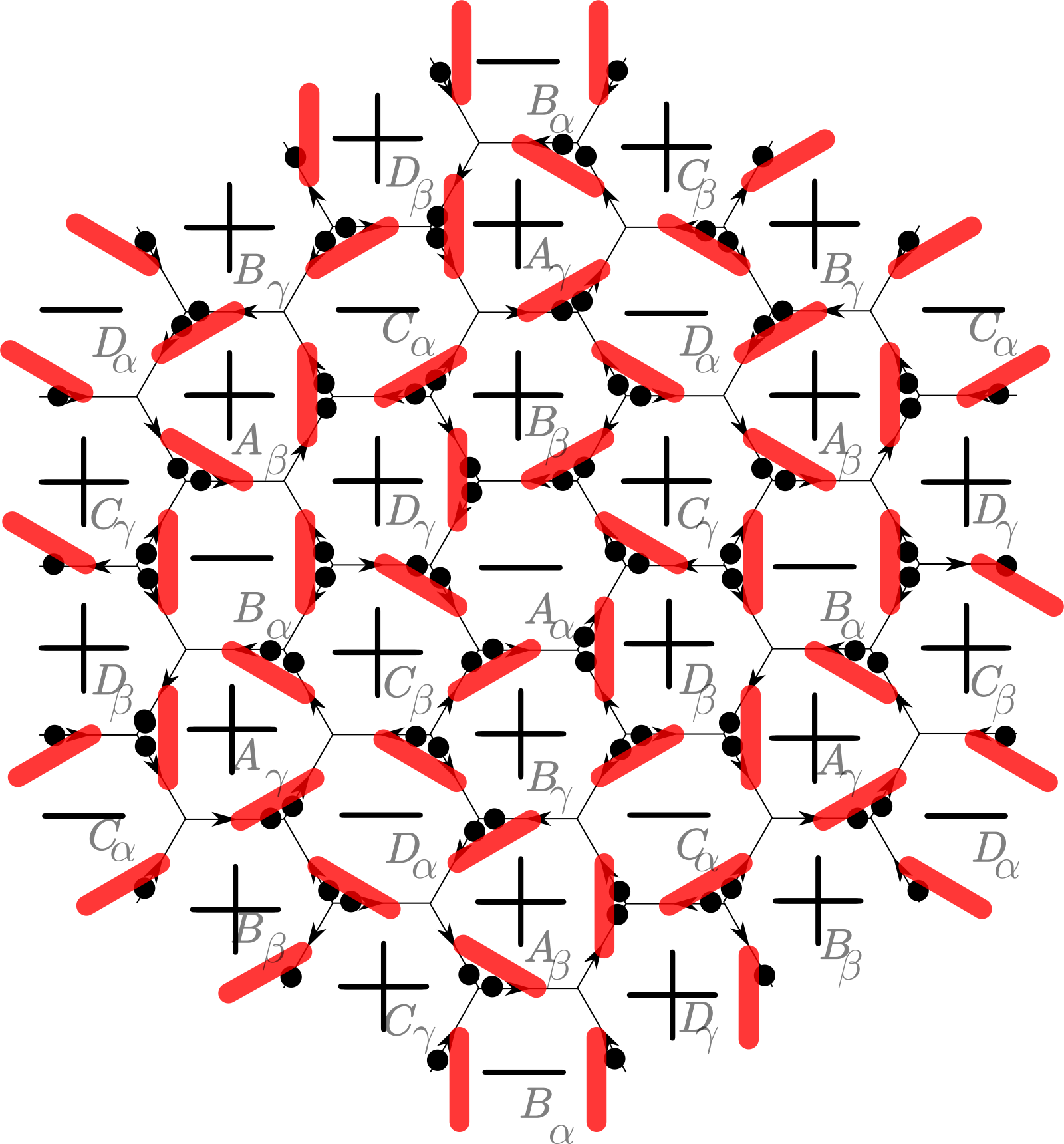}
  \end{center}
  \caption{\it Illustration of a pinwheel phase with 12 distinct sublattices with additional subscripts ($\alpha$, $\beta$, and $\gamma$). Here the pinwheels reside on $A_\alpha$ plaquettes, with the order parameters $M_{X_\alpha}=-M_{X_\beta}=-M_{X_\gamma}$ (cf.\ Subsection \ref{SUB:Height}), where $X$ is any of the sublattices $A$, $B$, $C$, or $D$.}
  \label{FIG:17Phase}
\end{figure}

Although we will not discuss this further, we have also explored the parameter space fixing $T_4 = T_8$ and varying $T_1 = T_7$ (while putting all other parameters to zero). For sufficiently large $T_1 = T_7$, the correlated pinwheels exist only on one third of the plaquettes of a given sublattice. As illustrated in Fig.\,\ref{FIG:17Phase}, one can then distinguish 12 sublattices $X_\alpha$, $X_\beta$, $X_\gamma$, where $X$ refers to any of the sublattices $A$, $B$, $C$, or $D$. Hence, translation invariance is now spontaneously broken in a different way. As one would expect, the transition that separates this phase from the previously discussed correlated pinwheel phase is first order.

\subsection{Height Model Representation of the $SU(2)$ Quantum Link Model}
\label{SUB:Height}

Because they will play an important role in our Monte Carlo algorithm, we now introduce $\Z(2)$-valued height variables that are associated with the four sublattices $A$, $B$, $C$, and $D$. In particular, the peculiar choice of link orientations, illustrated in Fig.\,\ref{Honeycomblattice}, facilitates the following definition of height variables. If the single rishon, which resides on a given link that separates two hexagons, is ahead of the orientation arrow located in the middle of that link, the height variables associated with the two adjacent hexagons are the same. On the other hand, if the rishon sits behind the orientation arrow, the corresponding height variables are different. If all height variables are changed simultaneously, the rishon configuration remains unchanged. Hence, the height variable representation is redundant. Up to this global redundancy, there is a one-to-one correspondence between the height variable configurations and the rishon configurations. In particular, the height variables guarantee that the Gauss law is automatically satisfied.

In the Monte Carlo calculations (to be discussed in the next section) we will also consider external non-Abelian charges in the $3/2$ representation of $SU(2)$. Obviously, the Gauss law is violated at the lattice sites where the charges are located. In order to incorporate configurations with external charges into the height variable description, we introduce a Dirac string that connects a pair of external $3/2$ charges. The rule for assigning height variables is reversed when a link that separates two adjacent hexagons is traversed by a Dirac string. Based on the resulting height variable configuration, the Gauss law is automatically violated at the two ends of a Dirac string, i.e.\ at the location of the external charges. The Dirac string itself can be deformed arbitrarily (with its ends fixed) without changing the physics.

In order to distinguish the various pinwheel phases, we now introduce four order parameters $M_A$, $M_B$, $M_C$, and $M_D$, using the height variables that are associated with the four sublattices $A$, $B$, $C$, and $D$. The order parameter $M_X$ is defined as the sum of all height variables $\pm 1$ on sublattice $X$. In a correlated pinwheel phase, all four sublattices order, i.e.\ $\langle M_X \rangle \neq 0$ for all $X$. The order parameter signatures for the 8 realizations of the correlated pinwheel phase are summarized in Table \ref{TAB:PinwheelPhases}.
\begin{table}
	\begin{center}
		\begin{tabular}{cccc|c}
			$M_A$	& $M_B$	& $M_C$	& $M_D$	& Pinwheel \\\hline
			$+$  	& $+$ 	& $+$ 	& $+$ 	& $A\circlearrowleft$\\
			$-$  	& $+$ 	& $+$ 	& $+$ 	& $A\circlearrowright$\\
			$-$  	& $+$ 	& $-$ 	& $+$ 	& $B\circlearrowleft$ \\	
			$-$  	& $-$ 	& $-$ 	& $+$ 	& $B\circlearrowright$\\
			$-$   	& $+$ 	& $+$ 	& $-$ 	& $C\circlearrowleft$ \\				$-$  	& $+$ 	& $-$ 	& $-$ 	& $C\circlearrowright$\\				$-$  	& $-$ 	& $+$ 	& $+$ 	& $D\circlearrowleft$ \\	
			$-$  	& $-$  	& $+$	& $-$ 	& $D\circlearrowright$\\
		\end{tabular}
	\end{center}
	\caption[Correlated pinwheel phase signatures.]{Order parameter signatures for the 8 correlated pinwheel phases. The arrow denotes the orientation of the pinwheels. The height variables are affected by an overall sign redundancy, not included here.}
	\label{TAB:PinwheelPhases}
\end{table}
In the uncorrelated pinwheel phase, on the other hand, the sublattice $X$, on which the pinwheels reside, does not order, i.e.\ $\langle M_X \rangle = 0$, but the other three sublattices still order.  The order parameter signatures for the 4 realizations of the uncorrelated pinwheel phase are summarized in Table \ref{TAB:UnCorrPinwheelPhases}.
\begin{table}
	\begin{center}
		\begin{tabular}{cccc|c}
			$M_A$	& $M_B$	& $M_C$	& $M_D$	& Pinwheel \\\hline
			$0$  	& $+$ 	& $+$ 	& $+$ 	& $A$ \\
			$-$  	& $0$ 	& $-$ 	& $+$ 	& $B$ \\
			$-$  	& $+$ 	& $0$ 	& $-$ 	& $C$ \\
			$-$  	& $-$ 	& $+$ 	& $0$ 	& $D$ 
		\end{tabular}
	\end{center}
	\caption[Uncorrelated pinwheel phase signatures.]{Order parameter signatures for the 4 uncorrelated pinwheel phases. The height variables are affected by an overall sign ambiguity, not included here.}
\label{TAB:UnCorrPinwheelPhases}	
\end{table}

\subsection{Fractionalization of $\Z(2)$ Flux Strings}
\label{SUB:FractionalFlux}

As discussed before, in contrast to the $\{5\}$-re\-pre\-sen\-ta\-ti\-on case, in the $\{4\}$-representation case it is not possible to localize the $\Z(2)$ center electric flux. As we will see in our Monte Carlo simulations, a $\Z(2)$ flux string connecting half-integer external charges then fractionalizes into two strands. Indeed the flux strands represent interfaces separating distinct realizations of pinwheel phases. The interface tension then manifests itself as a string tension. In the simulations we identify the flux strands by their energy density. Since the $\Z(2)$ flux is delocalized, it is not obvious whether only the energy or also the flux itself fractionalizes. Indeed, it seems difficult to imagine that something as elementary as a single unit of $\Z(2)$ flux could be divided into two halves. However, this is exactly what happens in the $\{4\}$-representation case.

In Subsection \ref{SUBSEC:ElecFlux} we showed how to identify center electric flux that flows through a closed loop (the vertical dotted lines in Fig.\,\ref{Flux}) wrapping around the periodic boundary. If the total number of rishons residing on the left side of the links intersected by the loop is odd, there is one unit of $\Z(2)$ flux flowing through the loop. If the number is even, on the other hand, there is no net $\Z(2)$ flux. When we count the rishons on the right side of the intersected links, we obtain the same result, at least if the extent of the lattice is even.

\begin{figure}
	\begin{center}
		\includegraphics[width=0.45\textwidth]{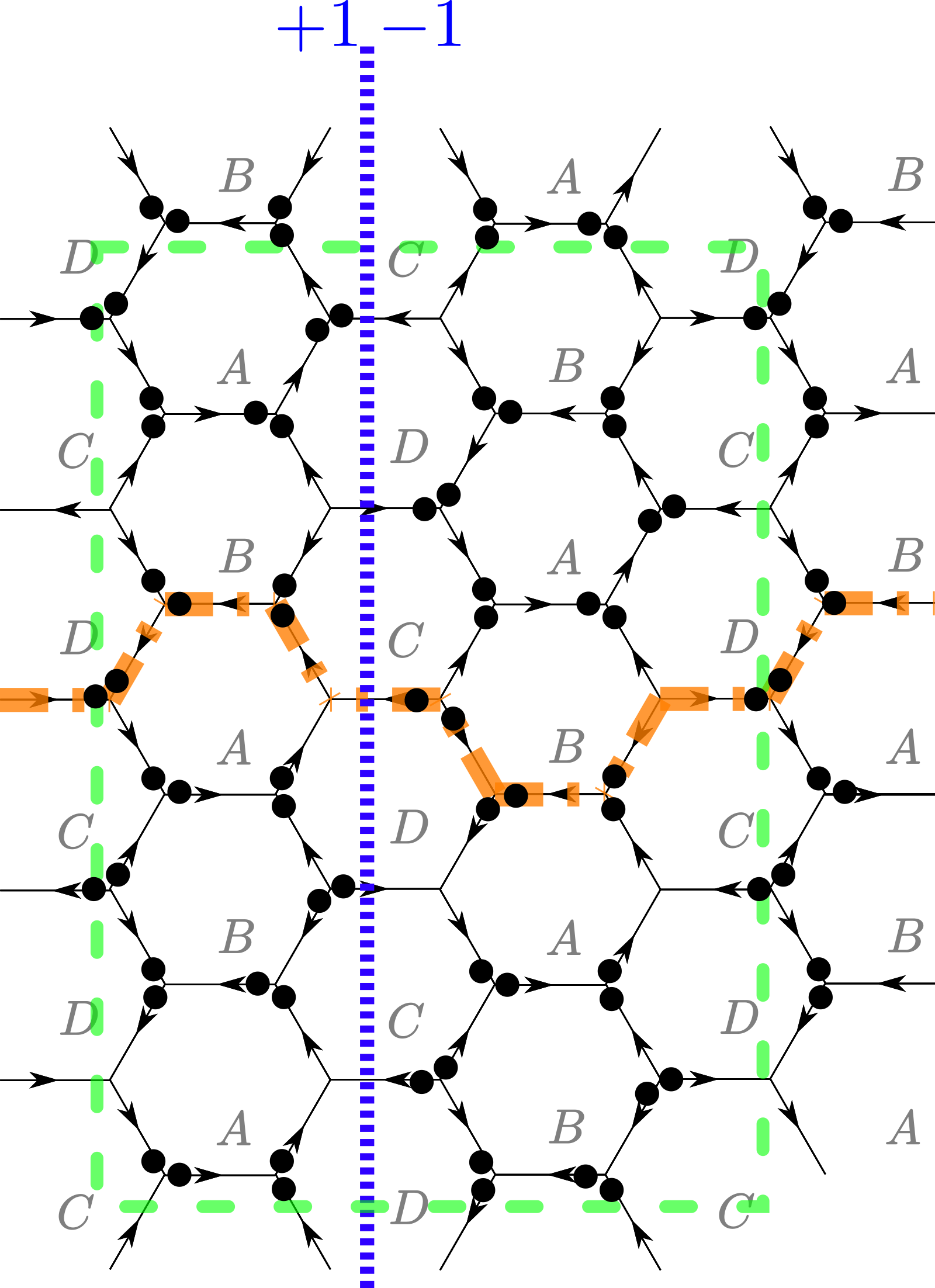}
	\end{center}
	\caption{\it An odd-sized lattice in a pinwheel phase, periodic over the green dashed box, always contains half a unit of $\Z(2)$ flux measured via a closed loop (vertical dotted line), together with a single associated domain wall (fat dashed links), separating two realizations of a pinwheel phase. Note that the sublattice structure is incommensurate with the odd-sized lattice.}
	\label{FIG:FluxOdd}
\end{figure}

In order to show that a single unit of $\Z(2)$ flux can indeed be divided into two halves, we now consider a periodic lattice with an odd extent, as illustrated in Fig.\,\ref{FIG:FluxOdd}. The wrapping loop now intersects an odd number of links, such that the difference of the number of rishons on the left and on the right is odd. This gives rise to even-odd or odd-even rishon-counts, which are indicated as $+1|-1$ or $-1|+1$. The fat dashed line in Fig.\,\ref{FIG:FluxOdd} represents a domain wall that separates two realizations of the pinwheel phase. Above the domain wall, the pinwheels reside on the A sublattice, and below the domain wall they reside on sublattice $B$. It should, however, be noted that the four-sublattice structure is incommensurate with the odd lattice extent. This is no problem. It just means that the odd extent forces at least one domain wall into the system. The fact that the rishon-count on the left and on the right side of the wrapping loop produces only one odd result shows that indeed one half of a $\Z(2)$ flux has been trapped in this volume of odd extent. As before, the fractionalized flux is still not localizable on individual links, but it is clear that it resides somewhere in the finite periodic volume.

It is interesting to note that there are two distinct types of half $\Z(2)$ flux states: those for which the rishon-count on the left of the wrapping loop is even and thus odd on the right ($+1|-1$), and those for which the count on the left is odd and thus even on the right ($-1|+1$). In Fig.\,\ref{FIG:FluxOdd} the two states are related to each other by shifting all rishons that reside on the domain wall from one end of the link to the other. This maintains Gauss' law but interchanges the rishon-count between even and odd, on both sides of the wrapping loop. It should be noted that the rishon-count depends on the choice of the location of the wrapping loop. In Fig.\,\ref{FIG:FluxOdd} the loop extends along the sublattices $C$ and $D$ and provides the rishon-count $+1|-1$. A shifted loop extending along the sublattices $A$ and $B$ would yield $-1|+1$. Still, what matters is that two distinct sectors exist.

To summarize, on a lattice of even extent a net flux is characterized by an odd-odd rishon-count ($-1|-1$ on the two sides of the wrapping loop), while the absence of net flux corresponds to an even-even count ($+1|+1$). On a lattice of odd extent, on the other hand, there is always one half of a $\Z(2)$ flux, characterized either by an even-odd ($+1|-1$) or by an odd-even ($-1|+1$) rishon-count. Hence, for lattices of both even and odd extents, there are two different $\Z(2)$ symmetry sectors, for each of the two spatial directions. The fact that a single unit of $\Z(2)$ flux can be divided into two halves is a unique feature of quantum link models, which does not arise in ordinary Wilson-type lattice gauge theories. It is possible only because in the $\{4\}$-representation a link carries a different $SU(2)$ representation on the left and on the right (one singlet and one doublet). When one uses the $\{5\}$-representation, on the other hand, the ends of a link carry the same $SU(2)$ representation (either both a singlet or both a doublet), cf.\ Fig.\,\ref{FIG:RishonLinkRep}. This situation also arises in Wilson's lattice gauge theory, except that the representation --- which is again the same on both ends of a link --- can be arbitrarily large.

\begin{figure}
	\begin{center}
		\includegraphics[width=0.45\textwidth]{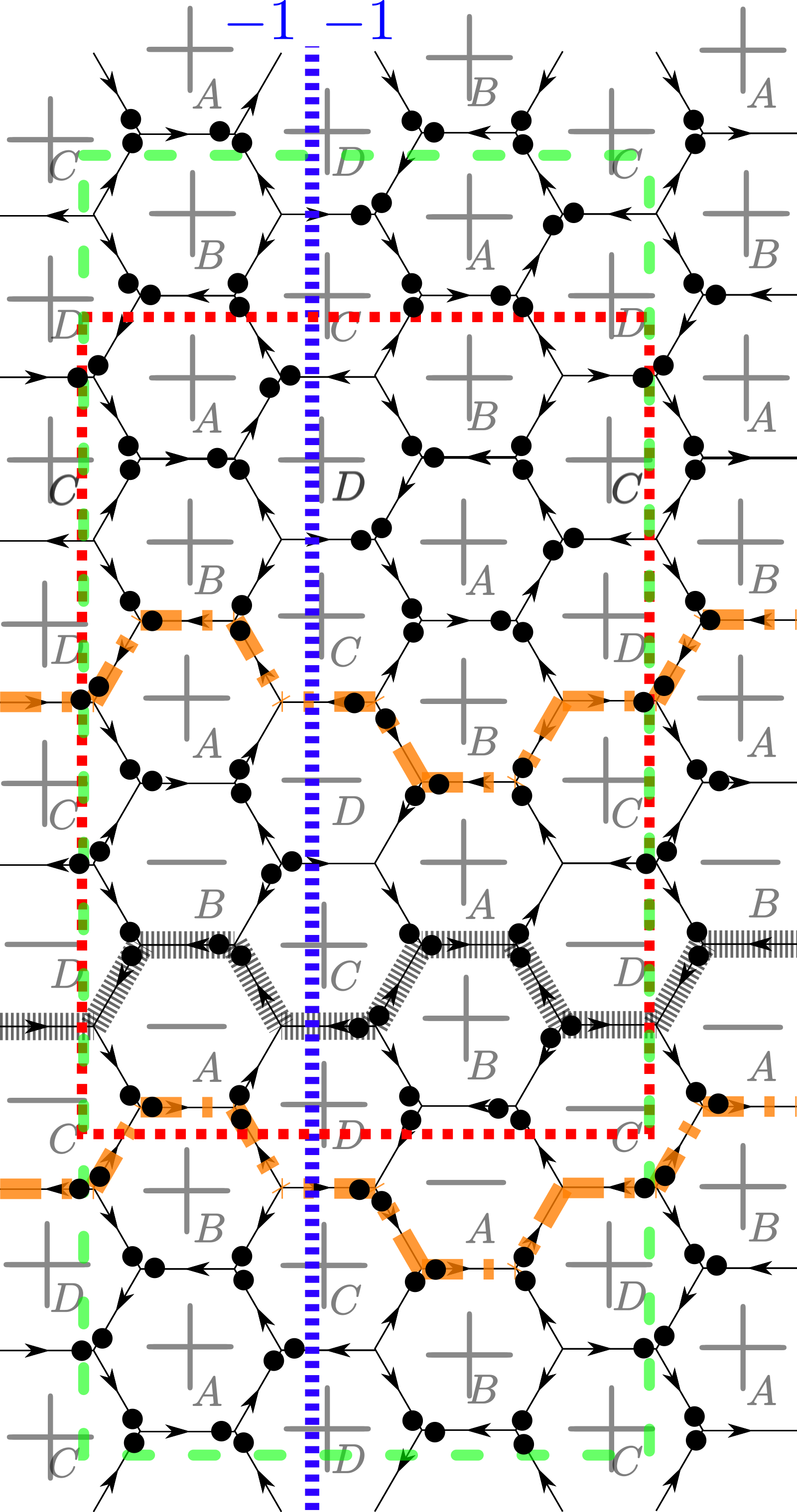}
	\end{center}
	\caption{\it On an even-sized lattice, the inclusion of a single Dirac string (hashed links) gives rise to two domain walls (fat dashed links), with one unit of $\Z(2)$ flux detected by the closed loop (vertical dotted line). The green dashed box is endowed with periodic boundary conditions. The configuration of Fig.\,\ref{FIG:FluxOdd} is included inside the red dotted box.}
	\label{FIG:FluxEven}
\end{figure}

In Fig.\,\ref{FIG:FluxEven} we have increased the lattice of Fig.\,\ref{FIG:FluxOdd} to be of even extent in the $y$-direction and we have enforced one unit of $\Z(2)$ flux to wrap around the $x$-direction by inserting a Dirac string (the hashed links). This gives rise to two domain walls (the fat dashed lines) that separate distinct correlated pinwheel phases. Although it is again impossible to localize the flux, it is natural to assume that it fractionalizes into two strands --- each carrying half of the $\Z(2)$ flux --- associated with the two domain walls. Indeed, the dotted line encircles a region that corresponds exactly to Fig.\,\ref{FIG:FluxOdd}, which contains half a unit of $\Z(2)$ flux. In Fig.\,\ref{FIG:FluxOdd} we did not insert any height variables, because a lattice of odd extent is incommensurate with the four sublattice structure. The even-extent lattice of Fig.\,\ref{FIG:FluxEven}, on the other hand, allows for the inclusion of height variables. The correlated pinwheel phase at the top and at the bottom of Fig.\,\ref{FIG:FluxEven} is characterized by the order parameter pattern $+++\,+$ for the four sublattices $A$, $B$, $C$, $D$. According to Table \ref{TAB:PinwheelPhases}, this phase has pinwheels on sublattice $A$, which are oriented counter-clockwise. The phase between the two domain walls, on the other hand, is characterized by the order parameter pattern $+-+\,-$. Note that below the Dirac string, this pattern changes to the equivalent pattern $-+-\,+$. According to Table \ref{TAB:PinwheelPhases}, this phase has pinwheels on sublattice $B$, which are again oriented counter-clockwise. As our numerical simulations will reveal, configurations in which the pinwheels are oriented in the same direction on both sides of a domain wall are energetically favored. In that case, two of the four order parameters $M_X$ change sign as we cross the domain wall.

\begin{figure}
	\begin{center}
		\includegraphics[width=0.45\textwidth]{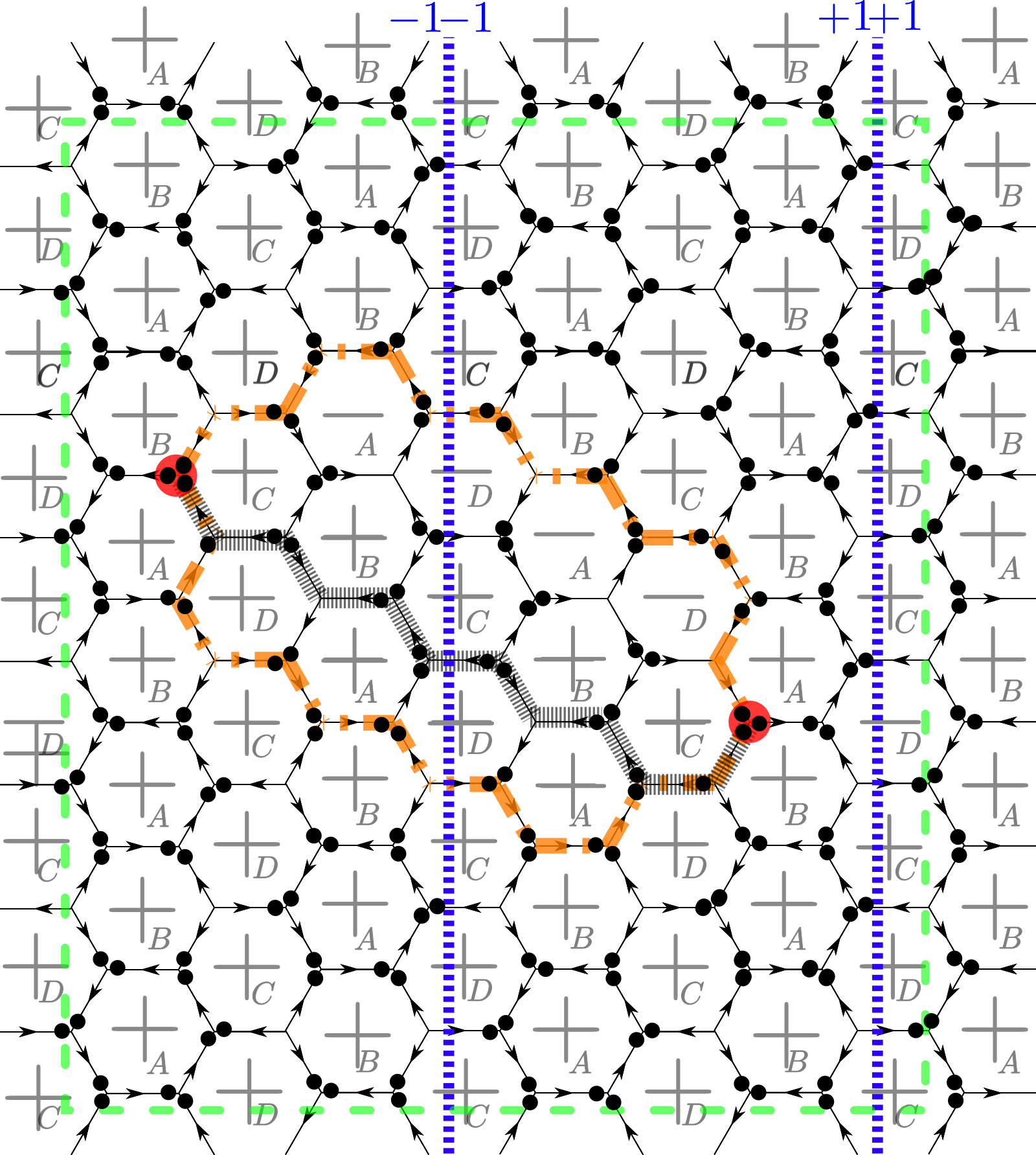}
	\end{center}
	\caption{\it The inclusion of two $3/2$ charges (red filled circles) connected by a Dirac string (hashed links) gives rise to two domain walls (fat dashed links) with a different pinwheel realization between the domain walls compared to the bulk (see Fig.\,\ref{FIG:FluxEven}). One unit of $\Z(2)$ flux is detected by a closed loop (vertical dotted line) between the charges, and no flux is present in the bulk (to the right of the charges). The green dashed box is endowed with periodic boundary conditions.}
	\label{FIG:FluxCharge}
\end{figure}

Fig.\,\ref{FIG:FluxCharge} illustrates a situation with two external $3/2$ charges in the background of a correlated pinwheel phase with order parameter pattern $++++$ (and thus with pinwheels on sublattice $A$ that are oriented counter-clockwise). As before, closed loops (the vertical dotted lines wrapping around the $y$-direction) are used to detect the $\Z(2)$ flux that flows through them. The loop that passes between the charges has an odd-odd rishon-count ($-1|-1$), thus indicating one unit of $\Z(2)$ flux. The other loop (to the right of both charges), on the other hand, yields an even-even rishon-count ($+1|+1$), and thus indicates the absence of flux. The situation is similar to Fig.\,\ref{FIG:FluxEven}. Again, there are two domain walls, however, they are no longer wrapping around the periodic $x$-direction, but instead end on the external charges. As before, the region between the domain walls is filled with a realization of the correlated pinwheel phase that is distinct from the one in the bulk, here with pinwheels on sublattice $C$, still oriented in the same counter-clockwise direction as in the bulk. In order to be able to define height variables, despite the fact that Gauss' law is violated at the location of the charges, they are connected by a Dirac string (across which the rules for assigning height variables are reversed). Above the Dirac string the order parameter pattern is $-++\,-$, and below it is $+--\,+$ (which is physically equivalent). Although, the $\Z(2)$ flux is delocalized, it is again natural to associate one half of it with each domain wall.

\section{Monte Carlo Simulations of the Height Model Representation}
\label{SEC:MonteCarlo}

In this section we present results of numerical simulations, concerning both the nature of the confining flux strings and the phase transition that separates correlated from uncorrelated pinwheel phases.

\subsection{Fractionalized Confining Strings in the Two Pinwheel Phases}

In Subsection \ref{SUB:FractionalFlux}, we have already familiarized ourselves with the properties of domain walls separating distinct pinwheel phases. In particular, they correspond to strands of confining strings carrying half a unit of fractionalized $\Z(2)$ center electric flux. This qualitative picture is confirmed quantitatively by our numerical simulations. Deep in the correlated pinwheel phase (for $T_4 = T_8$) we apply an efficient cluster algorithm that is described in Appendix \ref{APP:Cluster}. In the rest of the parameter space, where the cluster algorithm is not applicable, we employ a standard Metropolis algorithm. Although it is less efficient than the cluster algorithm, it is capable of extracting the physics of the model even close to its phase transition.

\begin{figure}
	\begin{center}
		\includegraphics[width=0.45\textwidth]{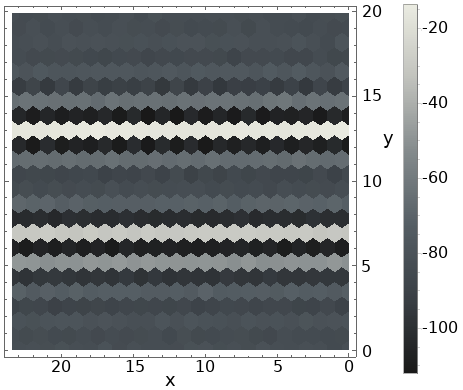}
	\end{center}
	\caption{\it Energy density on a $24\times24$ honeycomb lattice in the correlated pinwheel phase. A pair of domain walls is induced by an invisible Dirac string wrapping around the lattice in the $x$-direction. Here $\lambda = T_{4}/T_{8}=1$ and $\beta T_8=15$. The data are produced using the cluster algorithm.}
	\label{FIG:MonteCarloResSingleFluxCorr}
\end{figure}

Fig.\,\ref{FIG:MonteCarloResSingleFluxCorr} shows a pair of domain walls (representing one unit of $\Z(2)$ flux) wrapping around the periodic $x$-direction, deep in the correlated pinwheel phase at $T_4 = T_8$. The plot shows the energy density averaged over all Euclidean time-slices. Away from the domain walls (on the top, bottom, and in the middle of the plot) there are correlated pinwheel phases. The domain walls are characterized by rows of hexagonal plaquettes with alternating energies. Although this is not visible in the energy density itself, we have confirmed that the pinwheels on the two sides of a domain wall reside on different sublattices, but are oriented in the same direction. We have not investigated whether the domain walls correspond to rigid or rough interfaces (separating distinct pinwheel phases). Deep in the correlated pinwheel phase, it is plausible that the interfaces are rigid. Near the phase transition we expect them to be rough.

\begin{figure}
	\begin{center}
		\includegraphics[width=0.45\textwidth]{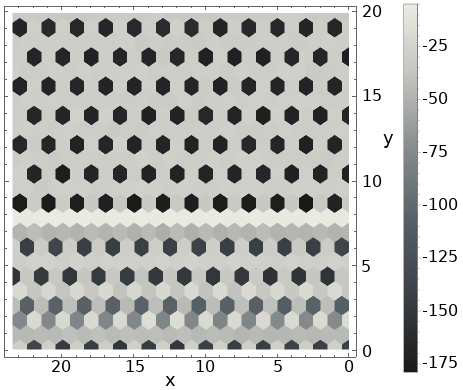}
	\end{center}
	\caption{\it Energy density on a $24\times24$ honeycomb lattice in the uncorrelated pinwheel phase. A pair of domain walls is induced by an invisible Dirac string wrapping around the lattice in the $x$-direction. Here $\lambda = T_{4}/T_{8}=\frac{1}{2}$ and $\beta T_8=15$. The data are produced using the Metropolis algorithm.}
	\label{FIG:MonteCarloResSingleFluxUnCorr}
\end{figure}

Fig.\,\ref{FIG:MonteCarloResSingleFluxUnCorr} illustrates the energy density corresponding to one unit of $\Z(2)$ flux that wraps around the periodic $x$-direction in the uncorrelated pinwheel phase. Similar to Fig.\,\ref{FIG:MonteCarloResSingleFluxCorr}, the flux fractionalizes into two strands that manifest themselves as a pair of domain walls. In contrast to Fig.\,\ref{FIG:MonteCarloResSingleFluxCorr} (which corresponds to $T_4 = T_8$), here the energy density explicitly shows the spontaneous breakdown of translation invariance. In the top part of Fig.\,\ref{FIG:MonteCarloResSingleFluxUnCorr} (above the upper domain wall) the pinwheels are on sublattice $C$, and between the domain walls they are on sublattice $B$.

\begin{figure}
	\begin{center}
		\includegraphics[width=0.45\textwidth]{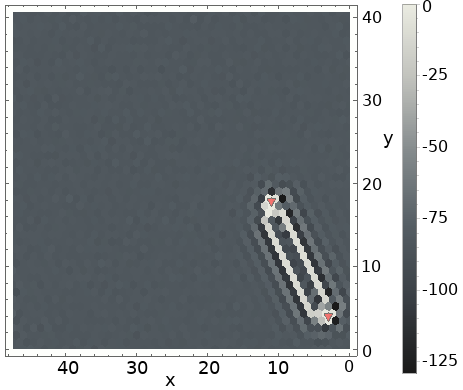}
	\end{center}
	\caption{\it Energy density for a $48\times48$ lattice in the correlated pinwheel phase with two external $3/2$ charges (red triangles) separated by 16 plaquettes. Here $\lambda = T_4/T_8 = 1$ and $\beta T_8 = 15$. The data are generated with the cluster algorithm.}	
	\label{FIG:MonteCarloResChargeCorr}
\end{figure}

\begin{figure}
	\begin{center}
		\includegraphics[width=0.47\textwidth]{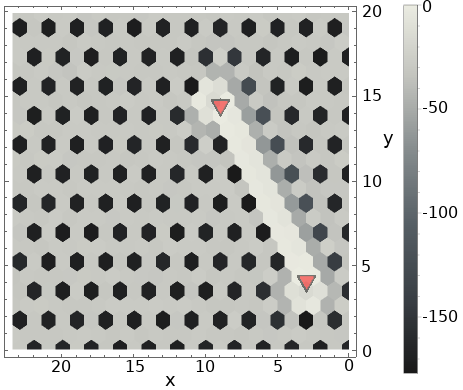}
	\end{center}
	\caption{\it Energy density for a $24\times24$ lattice in the uncorrelated pinwheel phase with two $3/2$ charges (red triangles) separated by 12 plaquettes. Here $\lambda = T_4/T_8 = \frac{1}{2}$ and $\beta T_8 =15$. The data are generated with the Metropolis algorithm.}	
	\label{FIG:MonteCarloResChargeUnCorr}
\end{figure}

We have also inserted pairs of external $3/2$ charges in order to investigate the confining strings that connect them. Fig.\,\ref{FIG:MonteCarloResChargeCorr} illustrates the situation in the correlated pinwheel phase, which resembles the ``cartoon'' Fig.\,\ref{FIG:FluxCharge}. In particular, one again sees that the flux fractionalizes into two strands that emanate from the charges. Although it is not obvious from the figure, we have verified that the bulk and the region between the flux strands are in distinct pinwheel phases. Fig.\,\ref{FIG:MonteCarloResChargeUnCorr} illustrates the corresponding situation in the uncorrelated pinwheel phase. In this particular case, deep in the uncorrelated pinwheel phase, the strands have coalesced to a single string.

\subsection{Phase Transition between two Crystalline Confining Phases}

Let us now investigate the nature of the phase transition that separates the correlated from the uncorrelated pinwheel phase. In both phases translation invariance is spontaneously broken. In the correlated pinwheel phase, in addition, reflection symmetry is spontaneously broken. This symmetry is restored in the uncorrelated pinwheel phase. If the phase transition is second order, the spontaneous breakdown of the $\Z(2)$ reflection symmetry suggests that it should be in the 3-d Ising universality class. Since translation invariance is spontaneously broken on both sides of the phase transition, this argument is not entirely straightforward.

\begin{figure}
	\begin{center}
		\includegraphics[width=0.47\textwidth]{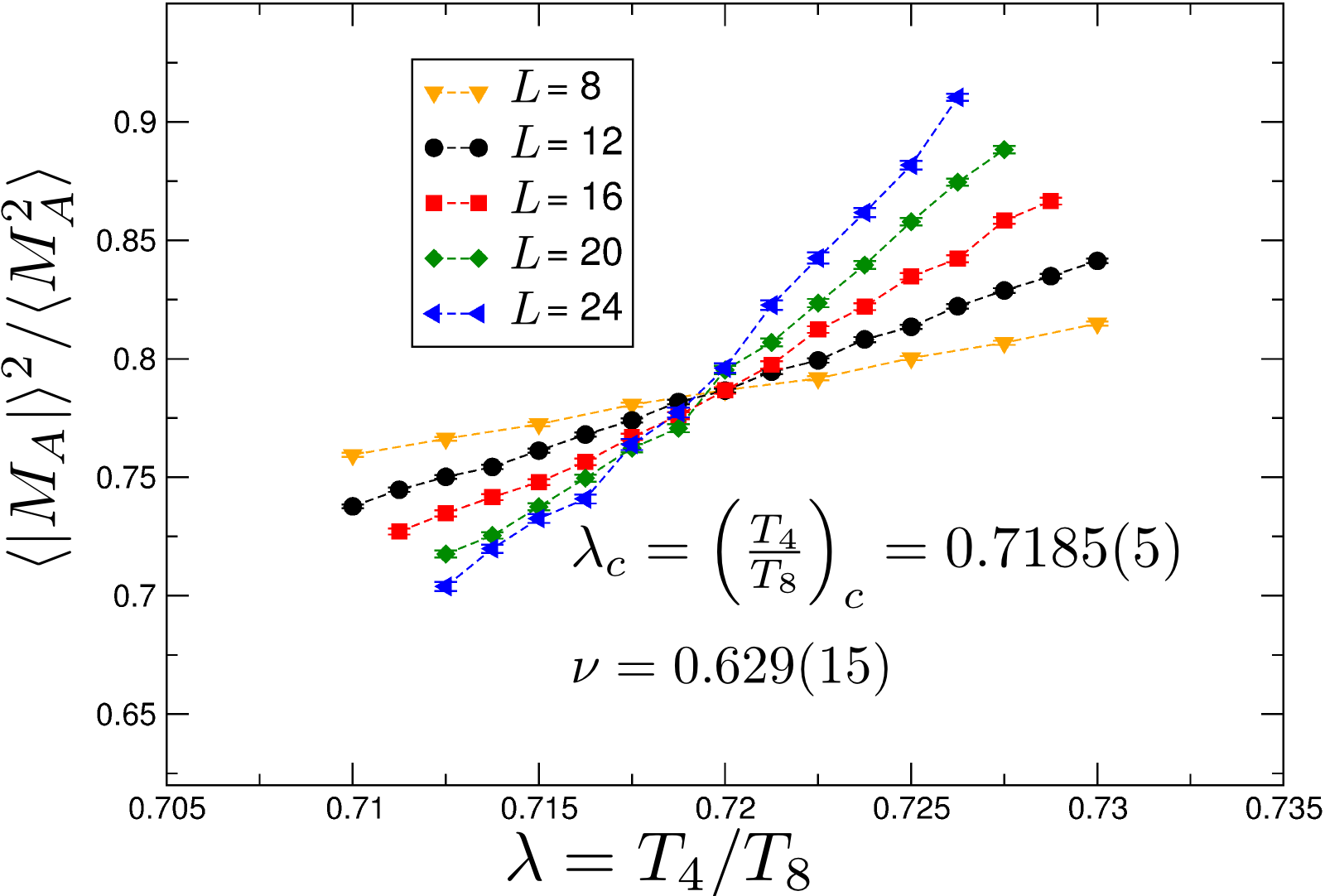}
	\end{center}
	\caption{\it Determination of the critical value $\lambda_c=(T_4/T_8)_c=0.7185(5)$ using the first Binder ratio $\langle|M_A|\rangle^2/\langle M_A^2\rangle$ for different lattice sizes $L$. The data are generated using the Metropolis algorithm with $\beta T_8 = \frac{3}{2} L$.}	
	\label{FIG:FirstBinderRation}
\end{figure}

\begin{figure}
	\begin{center}
		\includegraphics[width=0.47\textwidth]{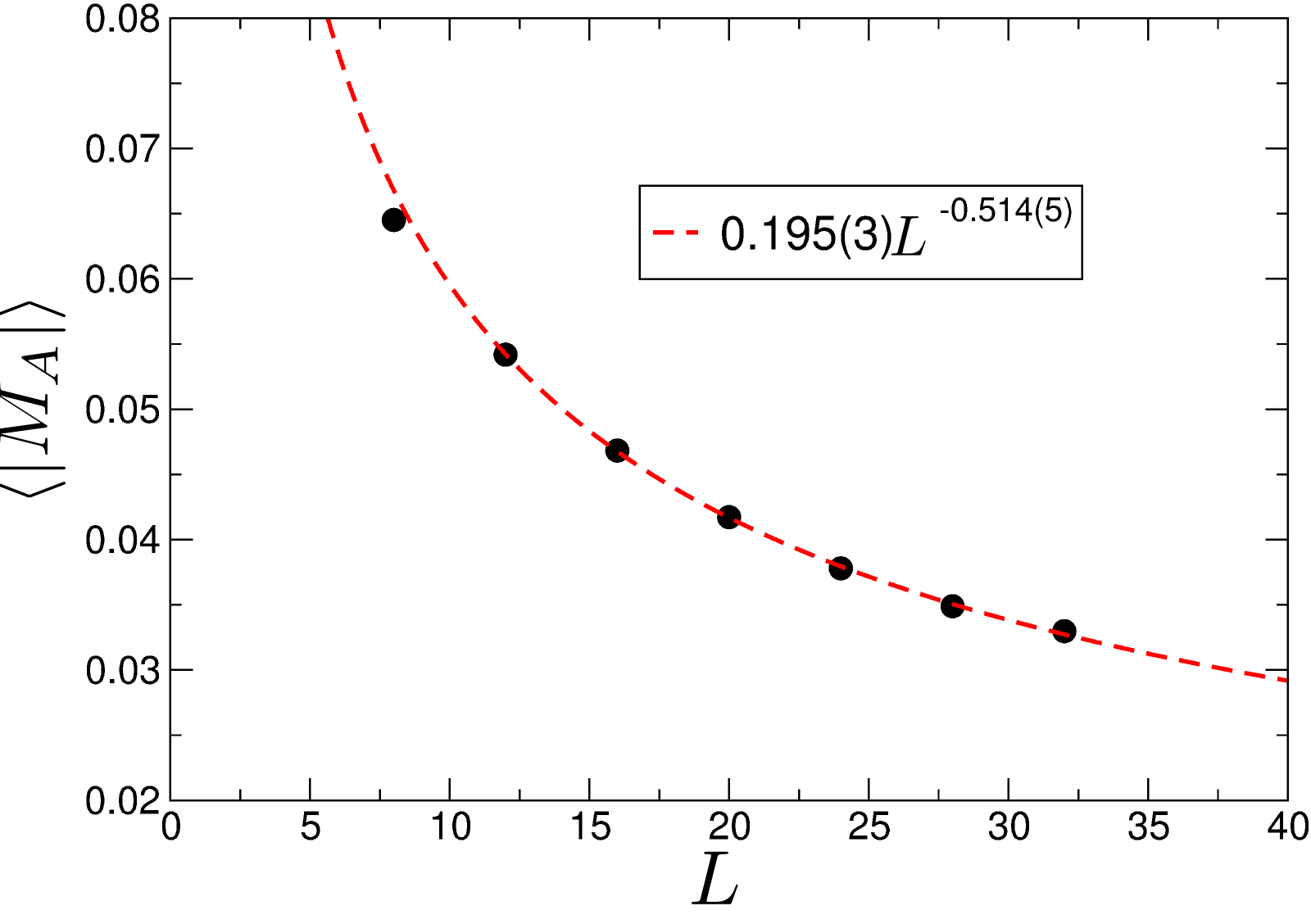}
	\end{center}
	\caption{\it Determination of the critical exponent $\beta$ from $\beta/\nu=0.514(5)$ by calculations of the magnetization $\langle |M_A| \rangle$ as a function of the lattice size $L$, at the critical point $\lambda_c=(T_4/T_8)_c=0.7185(5)$.}
	\label{FIG:CriticalBetaNu}
\end{figure}

Fig.\,\ref{FIG:FirstBinderRation} shows the finite-size scaling behavior of the first Binder ratio 
\begin{align}
\langle |M_A| \rangle^2/\langle M_A^2 \rangle & \propto f\left(\bar{\lambda}L^{1/\nu}\right),
\end{align}
where $f$ is a smooth function and $\bar{\lambda} = \frac{\lambda-\lambda_c}{\lambda_c}$. The fact that the various finite-volume curves intersect in one point indicates a second-order quantum phase transition at $\lambda_c = (T_4/T_8)_c = 0.7185(5)$. This value of the critical coupling $\lambda_c$ and the value of the critical exponent $\nu = 0.629(15)$ are extracted from a fit to the above finite-size scaling formula. The resulting value of $\nu$ is consistent with the known value $\nu = 0.629971(4)$ for the 3-d Ising universality class \cite{Kos2016}. Fig.\,\ref{FIG:CriticalBetaNu} shows $\langle |M_A| \rangle$ at the critical coupling as a function of the system-size. The expected finite-size scaling behavior
\begin{align}
\langle |M_A| \rangle & \propto L^{-{\beta}/{\nu}},
\end{align}
is again confirmed. The value of the critical exponent $\beta$ (not to be confused with the inverse temperature) follows from $\beta / \nu = 0.514(5)$. This is again consistent with the known value $\beta /\nu = 0.518149(3)$ for the 3-d Ising universality class \cite{Kos2016}.

\section{Conclusions}
\label{SEC:Conclusion}

We have investigated the $SU(2)$ quantum link model on the honeycomb lattice using the $\{4\}$-representation of the $SO(5)$ embedding algebra. Interestingly, this model is equivalent to the quantum dimer model on the Kagom\'e lattice. We constructed the most general gauge invariant single-plaquette ring-exchange Hamiltonian that respects the lattice symmetries. We then concentrated on the ring-exchanges of type 4 and 8 and investigated the phase diagram as a function of $\lambda = T_4/T_8$. We found a second-order quantum phase transition in the 3-d Ising universality class, which separates a correlated from an uncorrelated pinwheel phase. Both phases break lattice transition invariance spontaneously, with four emergent sublattices. In the correlated pinwheel phase, in which the orientation of the pinwheel dimers is coherent throughout the lattice, reflection symmetry is also spontaneously broken. This symmetry is restored in the uncorrelated pinwheel phase.

Although the phase transition itself is not unusual, the center electric $\Z(2)$ flux strings behave in a qualitatively new way. In contrast to Wilson-type lattice gauge theories, a string that carries a single unit of $\Z(2)$ flux fractionalizes into two separate strands. Although the flux is delocalized, we have constructed a finite system of odd extent that indeed traps half a unit of $\Z(2)$ flux which wraps around the periodic volume. Interestingly, the flux strands correspond to domain walls that separate different realizations of either the correlated or the uncorrelated pinwheel phase. Besides strings that wrap around the periodic volume, we have also investigated strings that connect a pair of external 3/2 charges. Again, the flux fractionalizes (unless one moves deep into the uncorrelated pinwheel phase). The region between the flux strands again corresponds to a different realization of a pinwheel phase compared to the bulk.

Our investigation demonstrates that quantum link models display qualitatively new phenomena that are absent in Wilson's lattice gauge theory. In particular, there are crystalline confined phases with spontaneously broken translation symmetry, which give rise to fractionalized electric flux strings. Crystalline confinement as well as the fractionalization of electric flux had already been observed in $(2+1)$-d $U(1)$ quantum link and quantum dimer models \cite{Ban13a,PhysRevB.94.115120}. This work extends these observations to the non-Abelian $SU(2)$ quantum link model. Although a single unit of $\Z(2)$ center electric flux may have seemed indivisible before, we have explicitly demonstrated that it can break up into two pieces. It is an interesting challenge to realize these intriguing quantum link dynamics in ultracold atom experiments.

\section*{Acknowledgment}

The research leading to these results has received funding from the Schweizerischer Na\-tio\-nal\-fonds and from the European Research Council under the European Union's Seventh Framework Programme (FP7/2007-2013)/ ERC grant agreement 339220. P.\ O.\ has been supported in part by a PSC-CUNY research award.

\appendix

\section{Cluster Algorithm}
\label{APP:Cluster}

\begin{figure}[ht!]
	\begin{center}
		\vspace{10pt}
		$\underset{P_{\rm{bond}}=1\hspace{25pt}P_{\rm{bond}}=1-\frac{e^{-\epsilon W_{E'}}\cosh(\epsilon T_{E'})}{e^{-\epsilon W_{E}}\cosh(\epsilon T_{E})}=1-\frac{e^{-\epsilon W_{2,6}}\cosh(\epsilon T_{2,6})}{e^{-\epsilon W_{3,5}}\cosh(\epsilon T_{{3,5}})}}{\includegraphics[width=0.45\textwidth]{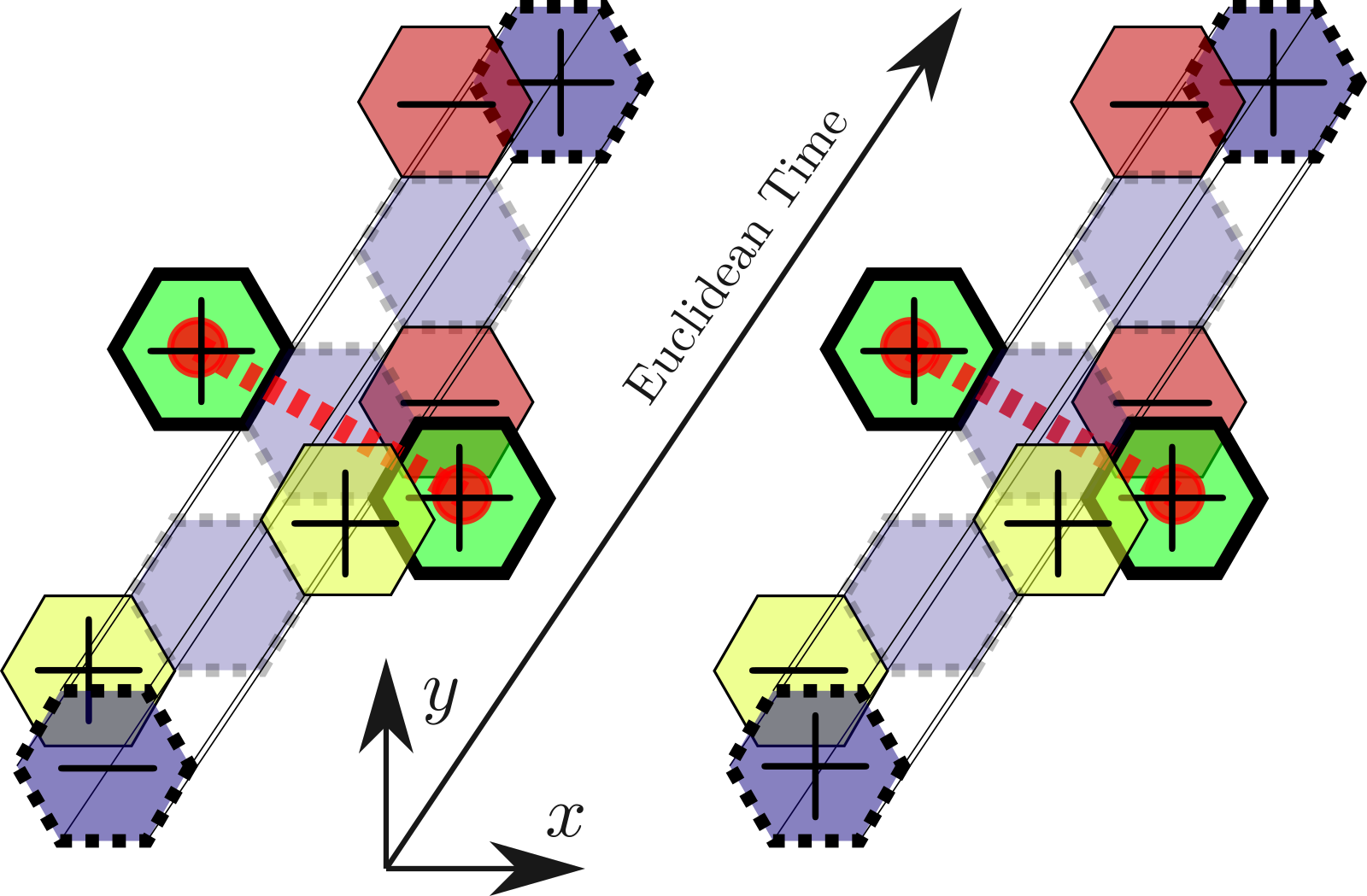}}$
	\end{center}
	\caption{\it A plaquette on sublattice $A$ (with dashed boundary) evolving over one discrete Euclidean time step. Pairs of $B$, $C$, and $D$ plaquettes in the three intervening Trotter-slices determine the corresponding Boltzmann weight. A spatial cluster-bond (fat dashed line) connects the height variables belonging to $C$ plaquettes (bold boundary) in the same Trotter-slice. Left: If the $A$ height variables undergo a transition (here from - to +), a spatial cluster-bond is placed with probability 1. Right: If the $A$ height variables remain constant in time, a spatial cluster-bond is put with a smaller probability.}
	\label{FIG:ClusterBondSpace}
\end{figure}

\begin{figure}[ht!]
	\begin{center}
		\vspace{10pt}
		$\underset{P_{\rm{bond}}=1\hspace{75pt}P_{\rm{bond}}=1-\frac{\sinh(\epsilon T_{4,8})}{\cosh(\epsilon T_{4,8})}}{\includegraphics[width=0.45\textwidth]{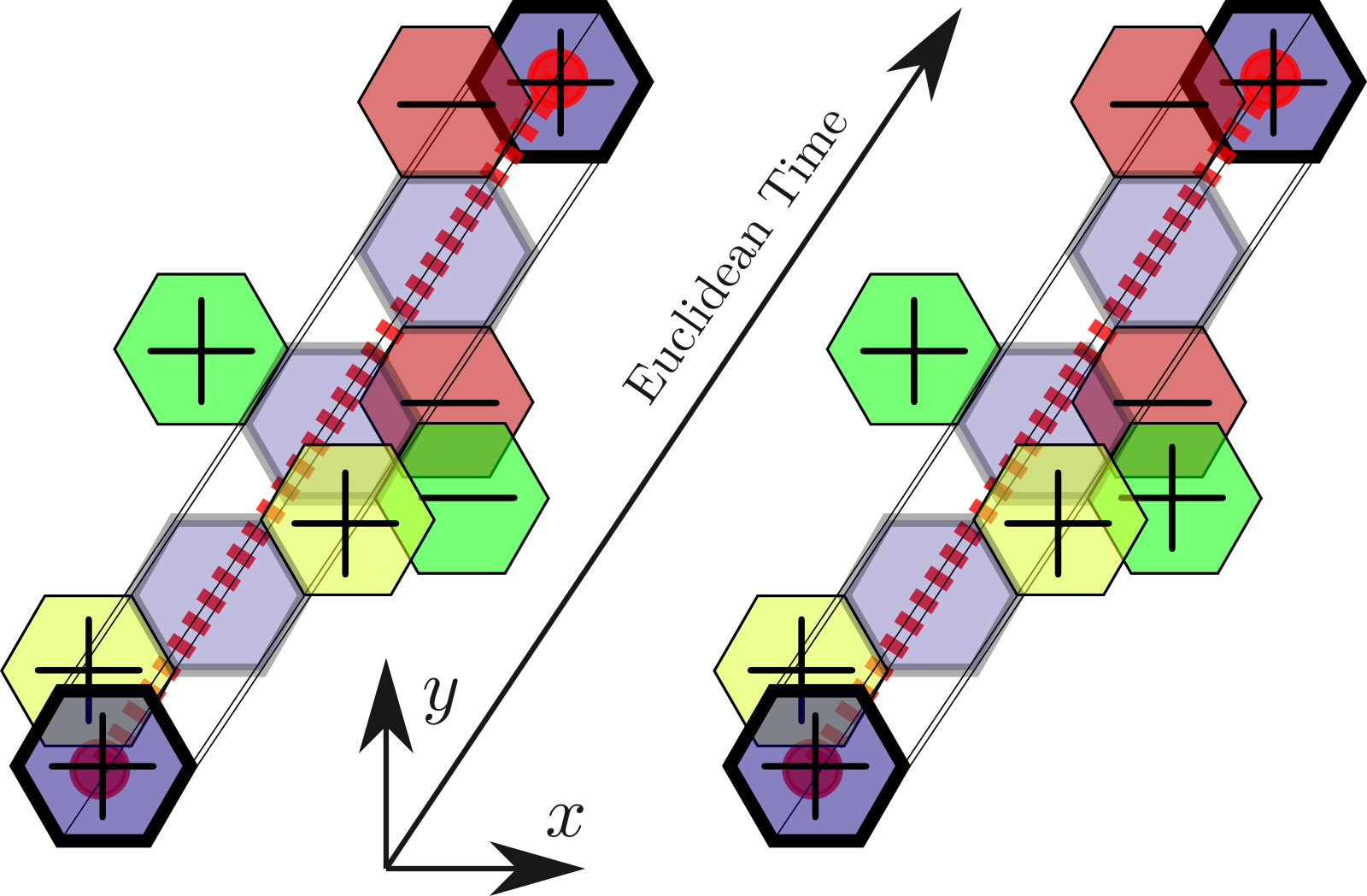}}$
	\end{center}
	\caption{\it A plaquette on sublattice $A$ (with bold boundary) evolving over one discrete Euclidean time step. Pairs of $B$, $C$, and $D$ plaquettes in the three intervening Trotter-slices determine the corresponding Boltzmann weight. A temporal cluster-bond (fat dashed line) connects the height variables belonging to $A$ plaquettes (bold boundary) separated by one Euclidean time step. Left: If the $B$, $C$, and $D$ height variables in the three intervening Trotter-slices are not all pairwise equal, a temporal cluster-bond is placed with probability 1. Right: If they are all pairwise equal, a temporal cluster-bond is put with a smaller probability.}
	\label{FIG:ClusterBondTime}
\end{figure}

Our numerical methods, either a standard Metropolis algorithm or a more efficient cluster algorithm, operate in the height variable representation. The cluster algorithm is applied in discrete Euclidean time, but has a continuous time limit, and could thus be implemented directly in the Euclidean time continuum. To ease the implementation, here we restrict ourselves to discrete time, working sufficiently close to the time-continuum limit. The Hamiltonian is decomposed into four pieces associated with the four sublattices, which results in a Trotter decomposition with four Trotter-slices per discrete Euclidean time step of size $\epsilon$. As a result, the height variables that are associated with different sublattices are residing in different Trotter-slices. This gives rise to an eight-height-variable interaction. Three pairs of height variables residing on three different sublattices (e.g.\ $B$, $C$, and $D$) in three intermediate Trotter-slices control the transition of a height variable on the fourth sublattice ($A$ in this case) over a single Euclidean time step. This is illustrated in Figs.\,\ref{FIG:ClusterBondSpace} and \ref{FIG:ClusterBondTime}.

We have constructed a multi-cluster algorithm that operates in the height representation. It updates the height variables on one sublattice at a time, while keeping the height variables on the other sublattices fixed. The algorithm identifies clusters of height variables. Only height variables of the same sign are put together in one cluster. This reflects the order of the correlated pinwheel phase. Once identified, the height variables in a single cluster are flipped with $50$ percent probability. The rules for assigning height variables to a cluster are described below. In order to allow for an overall flip of the height variables on an individual sublattice (which is essential for the efficiency of the cluster algorithm), we restrict ourselves to a region in parameter space with additional $\Z(2)$ symmetries. These additional symmetries are obtained by restricting the 18-dimensional parameter space to
\begin{align}
&W_1=W_7, & &W_2=W_6, & &W_3=W_5, & &W_4=W_8,\nonumber\\
&T_1=T_7, & &T_2=T_6, & &T_3=T_5, & &T_4=T_8.
\end{align}
In order to ensure the efficiency of the cluster algorithm, it must respect the correlated pinwheel order. This further restricts the parameters to
\begin{align}
W_{1,7}& \geq W_{2,6} \geq W_{3,5} \geq W_{4,8},\nonumber \\
T_{1,7}& = T_{2,6} = T_{3,5} = 0, \quad T_{4,8} \neq 0.\label{EQ:ClusterWeightConstraint}
\end{align}

Clusters are constructed by binding neighboring height variables on the same sublattice with cluster-bonds. Two height variables may be connected by a cluster-bond only if they have the same value. There are two types of cluster-bonds: those that connect spatial and those that connect temporal neighbors on the same sublattice. As illustrated in Fig.\,\ref{FIG:ClusterBondSpace}, spatial cluster-bonds (in this case on sublattice $C$) are put with probability $P_{\rm{bond}} = 1$ if the height variables (in this case on sublattice $A$) undergo a transition (here from - to +) in the corresponding Euclidean time step. This prevents the generation of forbidden configurations (with zero Boltzmann weight). On the other hand, if the height variables on sublattice $A$ are constant in time, the spatial cluster bond that connects height variables on sublattice $C$ is put with probability
\begin{align}
P_{\rm{bond}} &= 1-\frac{e^{-\epsilon W_{E'}}\cosh(\epsilon T_{E'})}{e^{-\epsilon W_{E}}\cosh(\epsilon T_{E})}.
\end{align}
Here $E$ is the current environment and $E'$ is the environment that results if a single height variable on sublattice $C$ changes sign. The same rules apply to the other intervening sublattices ($B$ and $D$ in this case). 

The rules for putting temporal cluster bonds between height variables (in this case on sublattice $A$) are illustrated in Fig.\,\ref{FIG:ClusterBondTime}. If the height variables in the intervening three Trotter-slices (residing on the sublattices $B$, $C$, and $D$) are not all pairwise equal, a temporal cluster-bond is put with probability $P_{\rm{bond}} = 1$. This again prevents the generation of forbidden configurations. On the other hand, if the height variables in the intervening three Trotter-slices are all pairwise equal, a temporal cluster-bond is put with probability
\begin{align}
P_{\rm{bond}} = 1 - \frac{\sinh(\epsilon T_{4,8})}{\cosh(\epsilon T_{4,8})}.
\end{align} 

\bibliographystyle{unsrt} 
\bibliography{kagomePRB_Bibliography}

\end{document}